\documentclass[aps,preprintnumbers,amsmath,amssymb]{revtex4}
\usepackage{graphicx}
\usepackage{dcolumn}% Align table columns on decimal point
\usepackage{bm}% bold math
\usepackage{graphicx}
\usepackage{epsfig}
\usepackage[usenames]{color}

\usepackage{setspace}
\usepackage[mathscr]{eucal}
\usepackage{amsthm}
\usepackage{float}

\begin{document}
\title{Two-dimensional structures  in the quintic Ginzburg-Landau equation}

 \author{Florent B\'erard, Charles-Julien Vandamme}
 \email{fberard@enseirb-matmeca.fr, cvandamme@enseirb-matmeca.fr}
 \affiliation{Department of Mathematics, ENSEIRB-MATMECA, Universit\'e Bordeaux 1, France 33405}

\author{Stefan C. Mancas}
\email{mancass@erau.edu}
\affiliation{Department of Mathematics, Embry-Riddle Aeronautical University, Daytona Beach, FL 32114-3900, USA}

\begin{abstract}
By using ZEUS cluster at Embry-Riddle Aeronautical University we perform extensive numerical simulations  based on a two-dimensional Fourier spectral method  Fourier spatial discretization and an explicit scheme for time differencing) to find the range of existence of the spatiotemporal solitons of the two-dimensional complex Ginzburg-Landau equation with cubic and quintic nonlinearities. We start from the parameters used by Akhmediev {\it et. al.} and slowly vary them one by one to determine the  regimes where solitons exist as stable/unstable structures. We present eight classes of dissipative solitons from which six are known (stationary, pulsating, vortex spinning, filament, exploding, creeping) and two are novel (creeping-vortex propellers and spinning ``bean-shaped'' solitons). By running lengthy simulations for the different parameters of the equation, we find ranges of existence of stable structures (stationary, pulsating, circular vortex spinning, organized exploding), and unstable structures (elliptic vortex spinning that leads to filament, disorganized exploding, creeping). Moreover, by varying even the two initial conditions together with vorticity, we find  a richer behavior in the form of creeping-vortex propellers, and spinning ``bean-shaped'' solitons.  Each class differentiates from the other by distinctive features of their energy evolution, shape of initial conditions, as well as domain of existence of parameters. 
\end{abstract}

\maketitle
\section{Introduction}

The complex cubic-quintic Ginzburg-Landau  equation (CCQGLE) is one of the most intensively studied equation describing weakly nonlinear phenomena in dissipative systems \cite{Aranson,Dodd,Saarloos,Hecke}. In fluid mechanics, it is also often referred to as the Newell--Whitehead equation after the authors derived it in the context of B\'enard convection \cite{Dodd,Drazin:1,Ablowitz}. Many basic properties of the equation and its solutions are reviewed in \cite{Aranson,Bowman,Brusch:1,Brusch:2}, together with applications to a vast variety of phenomena including nonlinear waves \cite{Aranson,Alvarez}, superconductivity \cite{Fabrizio,Bac}, Bose--Einstein condensation \cite{Kev1}, intra-pulse Raman scattering \cite{ivan}, liquid crystals and string theory. In particular, in nonlinear optics  it describes the pulse generation and signal transmission through an optical fiber \cite{7,Kaup:3}.

An important element  in the long time dynamics of pattern forming systems is a class of solutions which we call ``coherent structures" or solitons. A dissipative soliton is a self localized structure that can be a profile of light intensity, temperature or magnetic field, and is a  solution of a partial differential equation describing the evolution of a dissipative system. It is   localized and exists for an extended period of time and while it propgates its parts are experiencing gain/loss of energy  with the medium \cite{8,Doelman:1,Holmes}.   Whereas traditional solitons are stationary in time and preserve their shape upon interaction, some dissipative soliton solutions of the CCQGLE are non stationary. In Hamiltonian systems, stationary solitons exist as a result of a balance between diffraction/dispersion and nonlinearity \cite{Akhmediev:7}.  Diffraction spreads a beam while nonlinearity will focus it and make it narrower. The balance between the two results in stationary solitary wave solutions, which usually form a one parameter family. In dissipative systems with gain and loss, in order to have stationary solutions, the gain and loss must be also balanced \cite{Akhmediev:7, Akhmediev:3, Akhbook}.

 However, a dissipative system is known to be far from equilibrium and is defined by energy exchanges with external sources. Thus, there is no conserved quantity, which implies that they are not Hamiltonian systems. In dissipative systems, solitons are obtained by balancing gain and loss. As a result, whereas solutions are defined by one parameter family in Hamiltonian systems, in dissipative systems solutions are obtained  with varying amplitude and width that are fixed by the parameters of the equation. Therefore, the inverse scattering method \cite{Ablowitz} which is used to calculate solutions of integrable and some non integrable Hamiltonian systems cannot be used in our case.  Another relevant feature of dissipative systems is that they include energy exchange with external sources. Thus, energy can flow into the system through its boundaries. As long as the parameters in the system stay constant, the structure evolves by changing shape and exists indefinitely in time. The structure disappears when the source is switched off, or if the parameters are moved outside of the range of existence of the soliton solutions. 

Recent perturbative treatments based on expressions about the nonlinear Schr\"odinger equations are generalized to perturbations
of the cubic-quintic and derivative Schr\"odinger equations. The cubic Ginzburg-Landau admits a selected range of exact soliton
solutions \cite{Maruno}. Even the known special solutions have exotic dynamic responses to external conditions, so that sensitivity analysis of the parameters is paramount. These exist when certain relations between parameters are satisfied. However, this certainly does not imply that the equations are integrable. In reality, general dissipative nonlinear PDEs cannot be reduced to linear equations in any known way, therefore an insight to the type of solutions that the equation has may be based on numerical simulations \cite{Akhbook}. To investigate numerically the two-dimensional dissipative soliton solutions of the CCQGLE, and analyze their qualitative behavior, we used a powerful Fourier spectral method.

%%%%%%%%%%%%%%%%%%%%%%%%%%%%%%%%%%%%%%%%%%%%%%%%
\section {Numerical scheme of the CCQGLE}
 
In dimensionless form, the two-dimensional  CCQGLE with the corresponding cubic-quintic terms takes the form \cite{Ankiewicz:2,Akhmediev:7,Malo}
\begin{equation}\label{2.1}
\partial_tA=\epsilon A+(b_1+ic_1)\nabla_{\bot}^2 A-(b_3-ic_3) |A|^2A-(b_5-ic_5) |A|^4 A,
\end{equation}
where $t$ is the propagation distance, and the diffraction along the transverse plane with  Laplacian $\nabla_{\bot}^2 =\frac{\partial^2}{\partial x^2}+\frac{\partial^2}{\partial y^2}$ depends on both transverse coordinates $(x,y)$ \cite{Skarka2008,Akhmediev:3,Malo}. Originaly, the spatiotemporal structure  of the  CCQGLE was  introduced by \cite{Skarka2006,Holmes}, while the asymmetry between space and time was established in \cite{Skarka2007}, while the variational formulation for the one-dimensional solutions of the CCQGLE together with the factorization method  was established by \cite{Mancas:3,Rosu}. 

The physical parameters of the system are:  $\epsilon$ linear loss, $b_1$ angular spectral filtering, $c_1=0.5$ second-order diffraction coefficient, $b_3$ nonlinear gain/loss, $c_3=1$ self-focusing, $b_5$ saturation of the nonlinear gain/loss, and $c_5$ saturation of the nonlinear refractive index.

The equation  is solved by spectral methods, i.e., a Fourier spatial discretization and an explicit scheme for time differencing.  In addition, we analyze the dependence of both the shape and stability on the various parameters of the CCQGLE since the solutions experience interesting bifurcation sequences as the parameters are varied. 
  
The quantity that will be l monitored for each simulation is the energy $Q(t)=\int_{-\infty}^{\infty}\int_{-\infty}^{\infty} |A(x,y;t)|^2 \, dxdy$.
For a localized solution, $Q$ is finite and changes smoothly while the solution stays within the region of existence of the soliton. When $Q$ changes abruptly there is a bifurcation and the solution jumps from a branch of solitons that become unstable to another branch of stable solitons, or vice versa. As soon as the solution becomes unstable, $Q$ diverges until infinity or collapses to $0$. For a certain class of solutions, $Q$ will evolve periodically in some regime, and will converge to a finite value \cite{Crespo:1, Soto:2}.

%%%%%%%%%%%%%%%%%%%%%%%%%%%%%%%%%%%%%%%%%%%%%%%%
\subsection{Fourier Spectral Method}

The Fourier transform of a function $u(x,y)$ is defined by
\begin{equation}\label{four}
\mathcal{F}(u)(k_x,k_y)=\widehat{u}(k_x,k_y)=\frac{1}{2\pi}\int_{-\infty}^{\infty}\int_{-\infty}^{\infty}e^{-i(k_xx+k_yy)}u(x,y)\, dxdy,
\end {equation}
with the corresponding inverse
\begin{equation}
\mathcal{F}^{-1}(\widehat{u})(x,y)=u(x,y)=\frac{1}{2\pi}\int_{-\infty}^{\infty}\int_{-\infty}^{\infty}e^{i(k_xx+k_yy)}\widehat{u}(k_x,k_y)\, dk_xdk_y.
\end {equation}

The function $\widehat{u}(k_x,k_y)$ can be interpreted as the amplitude density of $u$ for wave numbers $k_x$, $k_y$. 

First, we write the Fourier transform of equation (\ref{2.1}) as
\begin{equation}\label{a}
\widehat{A_t}=\left[\epsilon-(b_1+ic_1)(k_x^2+k_y^2)\right]\widehat{A}-(b_3-ic_3)\widehat{|A|^2A }-(b_5-ic_5)\widehat{|A|^4A},
\end{equation}
and rearranging the terms in (\ref{a}), we consider the following ordinary differential equation in the Fourier space

\begin{equation}\label{b}
\widehat{A_t}=\alpha (k_x,k_y)\widehat{A}+\beta \widehat{|A|^2A }+\gamma \widehat{|A|^4A},
\end{equation}
with $\alpha (k_x,k_y)= \epsilon-(b_1+ic_1)(k_x^2+k_y^2)$, $\beta=-(b_3-ic_3)$ and $\gamma=-(b_5-ic_5)$.
In the Fourier space, (\ref{b}) contains the linear part $\alpha (k_x,k_y)\widehat{A}$, and the non-linear part $\beta \widehat{|A|^2A }+\gamma \widehat{|A|^4A}$. 

We solve the initial value problem for the above ODE  numerically using an explicit scheme  i.e., $4^\textrm{th}$ order Adams-Bashforth for the non-linear part and the exact solution for the linear part, which is given by
\begin{equation}\label{Exact}
\widehat{A}(t)=\widehat{A(x,y;0)}e^{\alpha (k_x,k_y)t}.
\end{equation}

%%%%%%%%%%%%%%%%%%%%%%%%%%%%%%%%%%%%%%%%%%%%%%%%%%%%%%%%%%%%%%%%%%%%%%%%
\subsubsection{Spatial discretization (Discrete Fourier Transform)}
\noindent
We use a standard discretization of a rectangular spatial domain $[-L/2,L/2] \times [-L/2,L/2]$ into $n \times n$ uniformly spaced grid points $X_{ij}=(x_i,y_j)$ with $\Delta x=\Delta y= L/n$ and $n$ even. Given $A(X_{ij})=A_{ij},\, i,j=1,2,\cdots,n$, we define the 2D Discrete Fourier transform (2DFT) of $A$ as
\begin{equation}\label{eq41}
\widehat{A}_{k_xk_y} = \Delta x \Delta y \sum_{i=1}^{n} \sum_{j=1}^{n} e^{-i(k_xx_i+k_yy_j)}A_{ij},\,\,\,\,\,\, k_x,k_y=-\frac{}{2}+1,\cdots, \frac{n}{2}
\end{equation}
and its inverse 2DFT as
\begin{equation}\label{eq5}
A_{ij} = \frac{1}{(2\pi)^2} \sum_{k_x=-n/2+1}^{n/2} \sum_{k_y=-n/2+1}^{n/2} e^{i(k_xx_i+k_yy_j)}\widehat{A}_{k_xk_y},\,\,\,\,\,\, i,j=1,2,\cdots, n.
\end{equation}
In (\ref{eq41})-(\ref{eq5}) the wavenumbers $k_x$ and $k_y$, and the spatial indexes $i$ and $j$,
take only integer values.

%%%%%%%%%%%%%%%%%%%%%%%%%%%%%%%%%%%%%%%%%%%%%%%%%%%%%%%%%%%%%%%%%%%%%%%%
\subsubsection {Temporal discretization ($4^\mathrm{th}$ order Adams-Bashforth)}

Given $t_\textrm{max}$ we discretize the time domain $[0,t_\textrm{max}]$ with equal time steps of width $\Delta t$ as  $t_n=n\Delta t , \, n=0,1,2,\cdots,$ and define
$\widehat{A}^n=\widehat{A}(t_n)$.
Initializing $\widehat{A}^n=\widehat{A}(t_n)$, we compute the nonlinear terms
$\mathcal{N}_3=\mathcal{F}\left(\left|\mathcal{F}^{-1}(\widehat{A}^n)\right|^2\mathcal{F}^{-1}(\widehat{A}^n)\right)$, and $\mathcal{N}_5=\mathcal{F}\left(\left|\mathcal{F}^{-1}(\widehat{A}^n)\right|^4\mathcal{F}^{-1}(\widehat{A}^n)\right)$
and advance the ODE (\ref{b}) in time
with time step $\Delta t$ using the 4-step Adams-Bashforth method. 

The discretized ODE has the following form
\begin{equation}
\widehat{A}^{n+1}=\widehat{A}^{n}e^{\alpha(k_x,k_y)t}+\frac{\Delta t}{24}\left[55f(\widehat{A}^{n})-59f(\widehat{A}^{n-1})+37f(\widehat{A}^{n-2})-9f(\widehat{A}^{n-3})\right],
\end{equation}
with exact solution of the linear part $\widehat{A}^{n+1}=\widehat{A}^{n}e^{\alpha(k_x,k_y)t}$,
and the nonlinear part $f$ defined by $f(\widehat{A})=\beta\mathcal{N}_1 +\gamma \mathcal{N}_2$.

%%%%%%%%%%%%%%%%%%%%%%%%%%%%%%%%%%%%%%%%%%%%%%%%%%%%%%%%%%%%%%%%%%%%%%%%
\subsection{Numerical implementation}\label{Numerical method}
\noindent
The language used is Fortran 90, with gfortran compiler. All the parameters needed by the program are written in a data file so that they can be easily modified, without systematic compilation. Among them, most important are the parameters of the  CCQGLE, which are shown in Table \ref{TabParam}.  
%For the ring solitons from \textsl{\S 4.4} we have used the 1D case of the parameters of \cite{Akhmediev:1,Soto:2,Ankiewicz:2}, while for the stationary and pulsating, the parameters are from \cite{Soto:2,Ankiewicz:2}. For the exploding soliton we have used the parameters from  \cite{Soto:3,Crespo:1}. All the parameters of the 1D case are extended in this paper to the 2D case.

An important subroutine is the Fast Fourier Transform (FFT) \cite{f90Recipes,Matlab}. This subroutine uses an algorithm that reduces computational time from $\mathcal{O}(n^2)$ to  $\mathcal{O}(n\log n)$ floating point operations. Moreover, it only works for periodic functions hence, caution must be made when we select the size of the domain for different cases of the soliton solutions. This must be chosen in such way that the solitons do not touch the boundaries, otherwise if the non-zero part of the solutions propagates through some boundary, it will spread throughout  the domain from the opposite side. 

This parallel computer (256 nodes dual Xeon 3.2 GHz processors, 1024 KB cache 4GB with Myrinet) at Embry-Riddle Aeronautical University is rated over 1 Tera-flop. Jobs were submitted thanks to Platform Lava installed on the cluster. This software manages to distribute all the jobs to different nodes, considering priorities. Once the output file are written, we used Tecplot 360 for visualization. The duration of computations depends on the equation parameters and mesh size. The most appropriate size to use was 512x512, but other types of meshes were considered. Very fast calculations took almost two hours, while longer ones took several days.
%%%%%%%%%%%%%%%%%%%%%%%%%%%%%%%%%%%%%%%%%%%%%%%%
\section{Numerical Simulations}
\subsection{Initial conditions}
\begin{figure}[htbp]
\centering
\includegraphics[width=350pt]{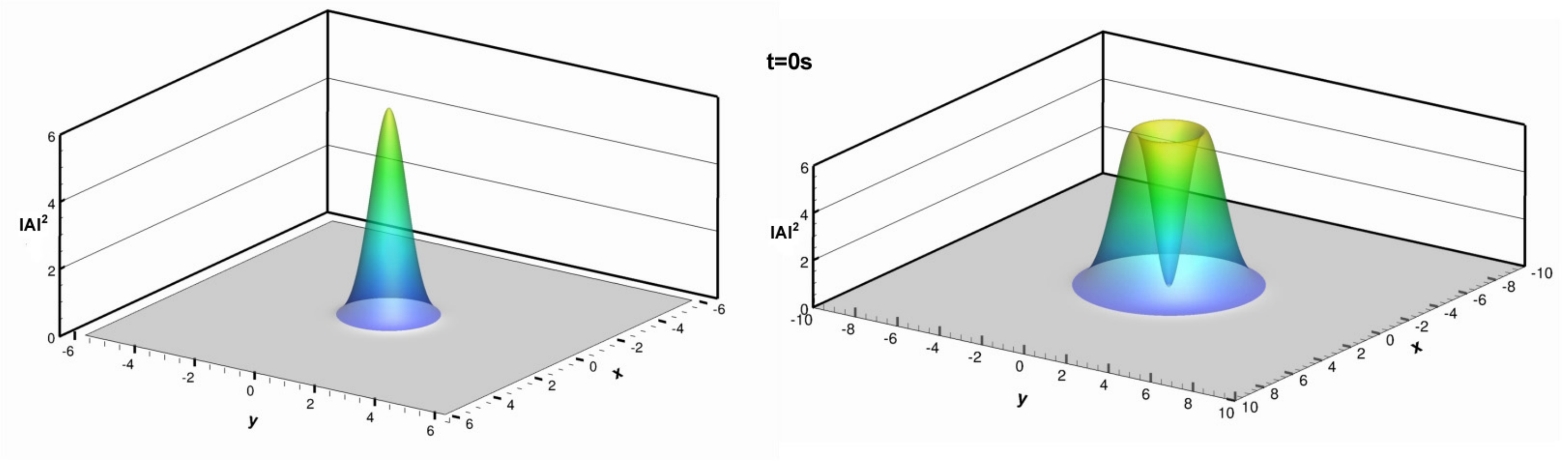}
\caption{Initial shapes of the field. Left: Gaussian shape. Right: Ring shape with vorticity $m=1$.}
\label{Initial}
\end{figure}

Localized structures with vorticity are obtained when initial conditions have radial symmetry, and hence we assume that the initial pulse can be reasonably well approximated by a bell-shaped, i.e.,
\begin{itemize}
\item[i.] Gaussian shape,
\begin{equation}
A(x,y;0)=A_0e^{-r^2}
\end{equation}
\item[ii.] Ring shape with rotating phase,
\begin{equation}
A(x,y;0)=A_0r^me^{-r^2}e^{im\theta}
\end{equation}
where $m$ is the degree of vorticity, $A_0$ is a real amplitude that should generate sufficient power to place the initial condition into a basin of attraction of a 2D soliton, and $\theta=\tan^{-1}{\big(\frac{\sigma_yy}{\sigma_xx}\big)}$ is the phase. The widths of each of the initial conditions could be either circular (radially symmetric) or elliptic, and are controlled by the parameters $\sigma_x$ and $\sigma_y$, with $r=\sqrt{(\sigma_xx)^2+(\sigma_yy)^2}$, see Fig. \ref{Initial}. In this paper we have used two vorticities $m=1, \, m=2$, but families of solitons of higher vorticity have been found before, see \cite{Soto:4}.
\end{itemize}

%%%%%%%%%%%%%%%%%%%%%%%%%%%%%%%%%%%%%%%%%%%%%%%%%%%%%%%%%%%%%%%%%%%%%%%%
\subsection{Parameters}

The existence of stable solutions is strongly linked to the parameters of the equation. One of the most important issue is to find the continuous range of existence of solitons of the same type in parameters space. Moreover, several solutions can coexist for the same set of parameters. These solitons are not necessarily stable. Hence, stability is controlled by the parameters of the equation and by the choice of the initial conditions.

Starting  from the parameters used by the Akhmediev\rq{}s group, namely rows 1,2 of Table \ref{TabParam} from \cite{Soto:2}, while rows 3,4,5 from \cite{Akhmediev:1}, for each class of solitons our goal was to find the regime of existence as stable structures. Hence,  we vary one by one all the  parameter in a range of width $0.2$ while keeping all the other parameters constant. When we are outside of the range of existence of the solitons, we return to the initial value, and vary the other remaining parameters.  After several simulations, we noticed that $b_3$ and $\epsilon$ seemed to be the more sensitive, i.e., small changes of $b_3$ or $\epsilon$ implied significant  changes of the solution. Thus, a step of $0.01$ has been chosen for these parameters whereas a step of $ 0.05$ has been used for $b_1$, $b_5$ and $c_5$. However, this behavior is not a general rule, so, in some cases, a smaller step has been used for the last three parameters. All the ranges are presented  in section \ref{sims}, when we study each class of solutions separately. 
\begin{table}
\begin{center}
\begin{tabular}{|c|ccccccc|}
\hline 2D solitons & $\epsilon$ & $b_1$ & $c_1$ & $b_3$ & $c_3$ & $b_5$ & $c_5$\\
\hline
stationary  &-0.045 & 0.04 & 0.5 & -0.211& 1 & 0.03 & -0.08  \\
\hline
pulsating  & -0.045 & 0.04 & 0.5 & 0.37 & 1 & 0.05 & -0.08 \\
\hline
spinning, filament & -0.1 & 0.1 & 0.5 & -0.8 & 1 & 0.04 & -0.02  \\
\hline
exploding & -0.1 & 0.125 & 0.5 & -1 & 1 & 0.1 & -0.6 \\
\hline
creeping, propeller & -0.1 & 0.101 & 0.5 & -1.3 & 1 & 0.3 & -0.101 \\
\hline
\end{tabular}
\end{center}
\caption{Initial sets of parameters from which we will start the simulations.}
\label{TabParam}
\end{table}

%%%%%%%%%%%%%%%%%%%%%%%%%%%%%%%%%%%%%%%%%%%%%%%%

\section{Simulation Results}\label{sims}
%%%%%%%%%%%AAAAAAAAAAAA%%%%%%%%%%%%%%%%%%%%%%%%%%%%%
\subsection{Stationary solitons}\label{Stationary}
\subsubsection{Description}

\begin{figure}[htbp]
\centering
\includegraphics[width=250pt]{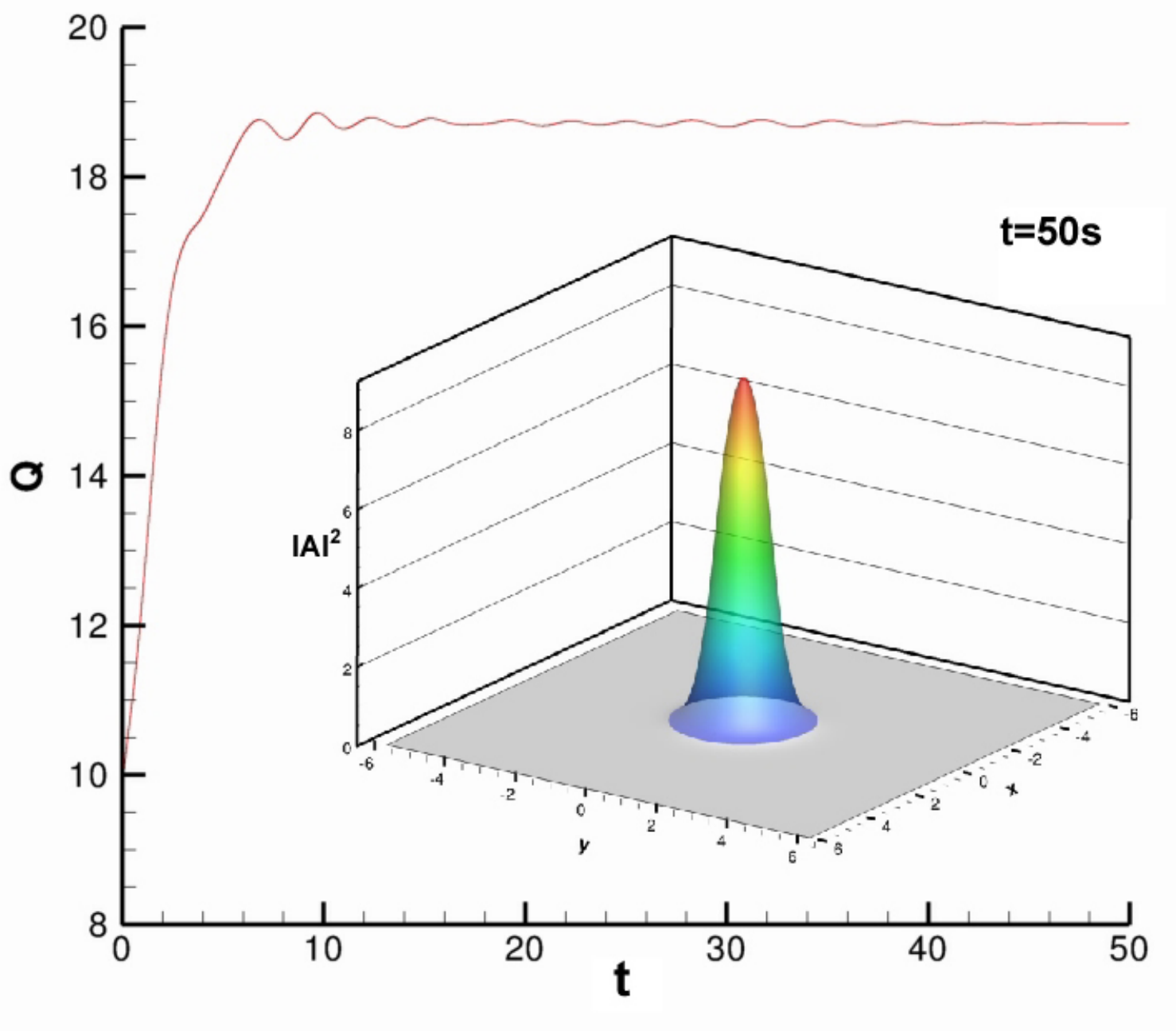}
\caption{Stationary soliton. Energy is concentrated in the center of the domain, in a bell-shaped structure. Parameters: row 1 of Table \ref{TabParam}.}
\label{Normal}
\end{figure}

For this class the initial shape is a circular 2D Gaussian with amplitude $A_0=2.5$, and widths $\sigma_x=\sigma_y=1$. For start-up  simulations we used row 1 of Table \ref{TabParam}. Steady solitons are the simplest solutions we can find, since after a transition period of stabilization, the shape of the solitary waves remains the same. The energy becomes 
constant after $40 \, s$. The soliton profile is always a Gaussian, and each profile
remains Gaussian, see Fig. \ref{Normal}.
Finally, energy converges to $Q(t)=18.6 \,s$, and stabilizes forever if the system's parameters are not changed.

\subsubsection{Ranges of parameters}
Varying parameters one by one, as explained before, the ranges of existence for the stationary solitons as stable structures are found and presented in Table \ref{TabStat}. Outside each interval, their energy either decays or diverges. 

\begin{table}
\begin{center}
\begin{tabular}{|c|c|}
\hline Parameters & Stationary \\
\hline
$b_1$  & [0.085, 0.235] \\
\hline
$\epsilon$  & [-0.245, -0.105]  \\
\hline
$b_3$ & [-0.35, -0.20]  \\
\hline
$b_5$  & [0.06, 0.18]  \\
\hline
$c_5$ & [-0.23, -0.11] \\
\hline
\end{tabular}
\end{center}
\caption{Parameter ranges of existence of stationary solitons.}
\label{TabStat}
\end{table}

%%%%%%%%%%%%%%%%BBBBBBBBBB!111111111111111111111111111%%%%%%%%%%%%%%%%

\subsection{Pulsating solitons}
\subsubsection{Description}

\begin{figure}[htbp]
\centering
\includegraphics[width=350pt]{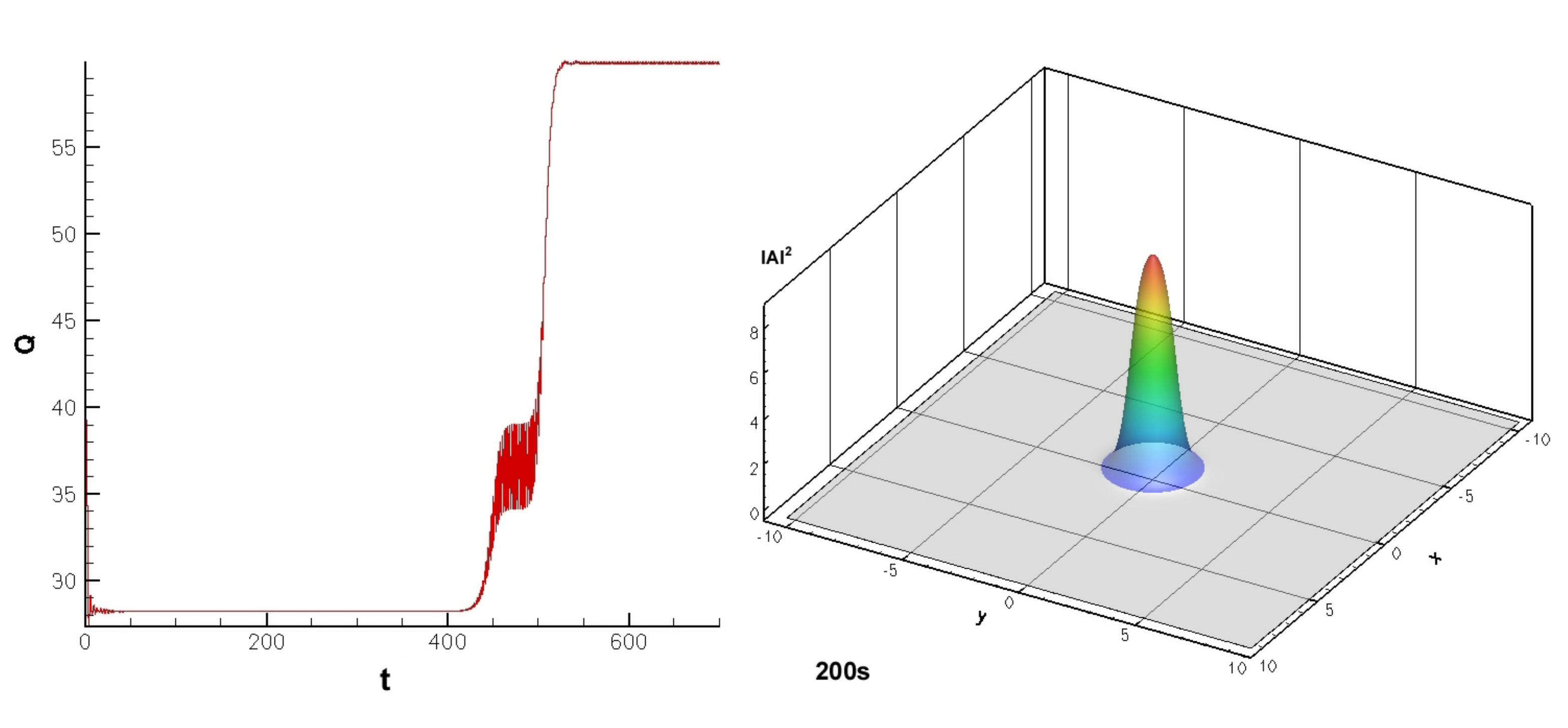}
\caption{Pulsating soliton with changing state. The energy, on the left shows these transitions. Right: bell-shaped soliton $t=200s$.Parameters: row 3 of Table \ref{TabParam}.}
\label{Pulse1}
\end{figure}

Many authors have found pulsating solitons especially in 1D case \cite{Akhmediev:1,Crespo:1,Mancas:6}, and in 2D \cite{Akhmediev:4,Soto:2,Ankiewicz:2}. The following results  contain two interesting features: pulsations and transitions between stable and unstable states. As it was explained  in \cite{Akhmediev:2}, a soliton with a very interesting pulsating behavior was discovered for a slightly asymmetric   initial shape which  is also Gaussian with amplitude $A_0=5$, and widths  $\sigma_x=0.8333$ and $\sigma_y=0.9091$, see Fig. \ref{Pulse1},  for the set of parameters of row 2 of Table \ref{TabParam}.

\begin{figure}
\centerline{\includegraphics[width=12cm]{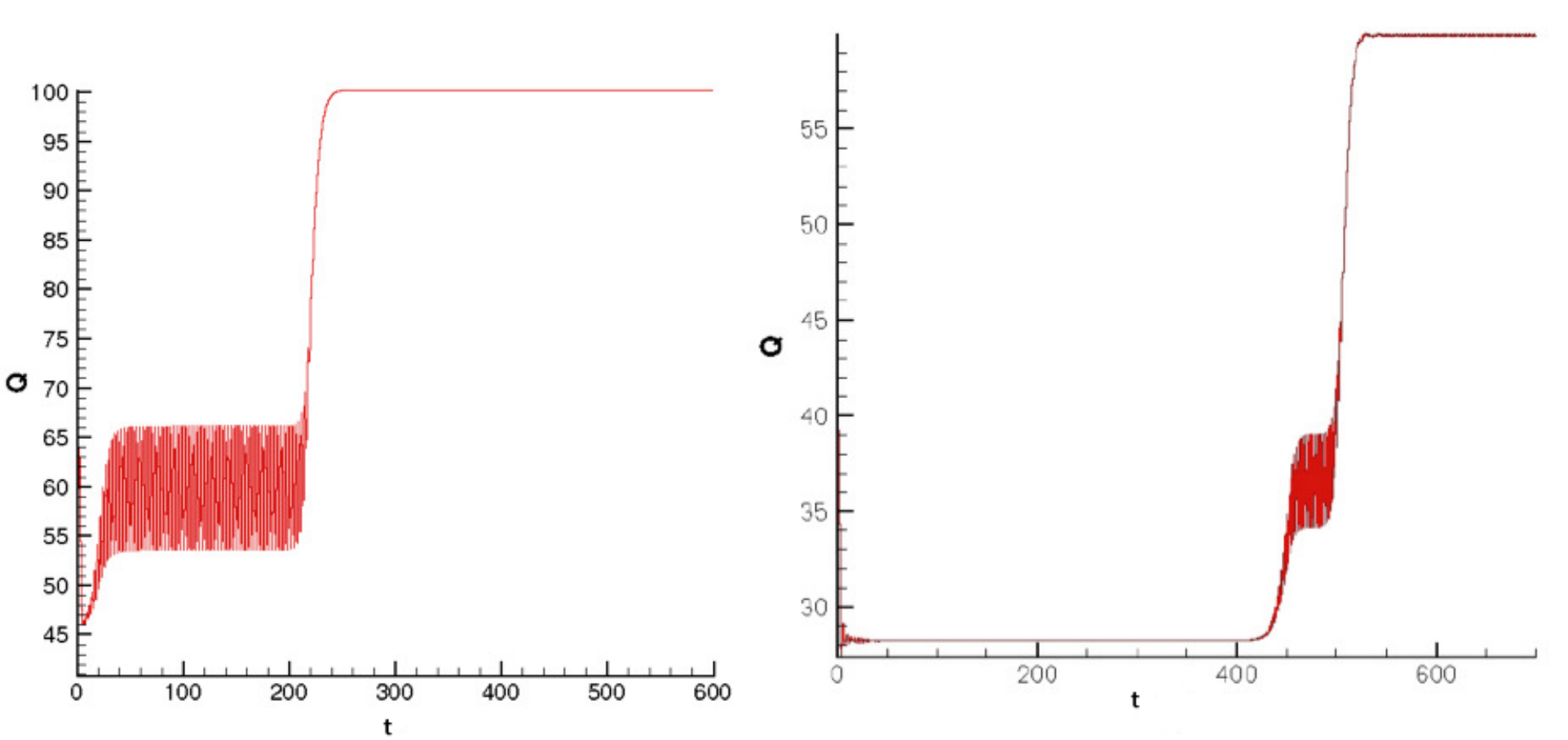}}
\caption{Energy of pulsating solitons. Left: $b_3=0.40$. Right: $b_3=0.37$.}
\label{puls3}
\end{figure}

For short simulations up to $400 \,s$, one might think that in this case the solitons evolve as a stationary ones, i.e. the energy $Q$ does not transition between stable/unstable states and oscillate. After longer simulations we found that the  energy increases and takes a pulsating behavior before increasing again to a new stationary state. During the pulsating period, the soliton keeps a radially symmetric profile at the maximum of energy, whereas for the minimum enerhy, the beam elongates alternatively in different directions. However, other pulsating solitons exist and do not evolve in the same way. For example, for two different values of $b_3$, the energy evolves differently, see Fig. \ref{puls3}. With $b_3=0.37$, after a short pulsating behavior, the soliton reaches a stationary stable state around $400\,s$, then  a  perturbation makes the solitons evolve towards a higher state with a new pulsating energy. This pulsating behavior lasts around $100\,s$ (right panel of Fig. \ref{puls3}).  For $b_3=0.40$, the first stationary state does not appear, so the energy starts oscillating really fast right from the beginning. This pulsating state lasts about $200\,s$ (left panel of Fig. \ref{puls3}). Furthermore, the energy values where both solitons converge after the pulsating state are  different ($Q= 60$ for $b_3=0.37$ and $Q= 100$ for $b_3=0.40$).

\begin{figure}[htbp]
\centering
\includegraphics[width=250pt]{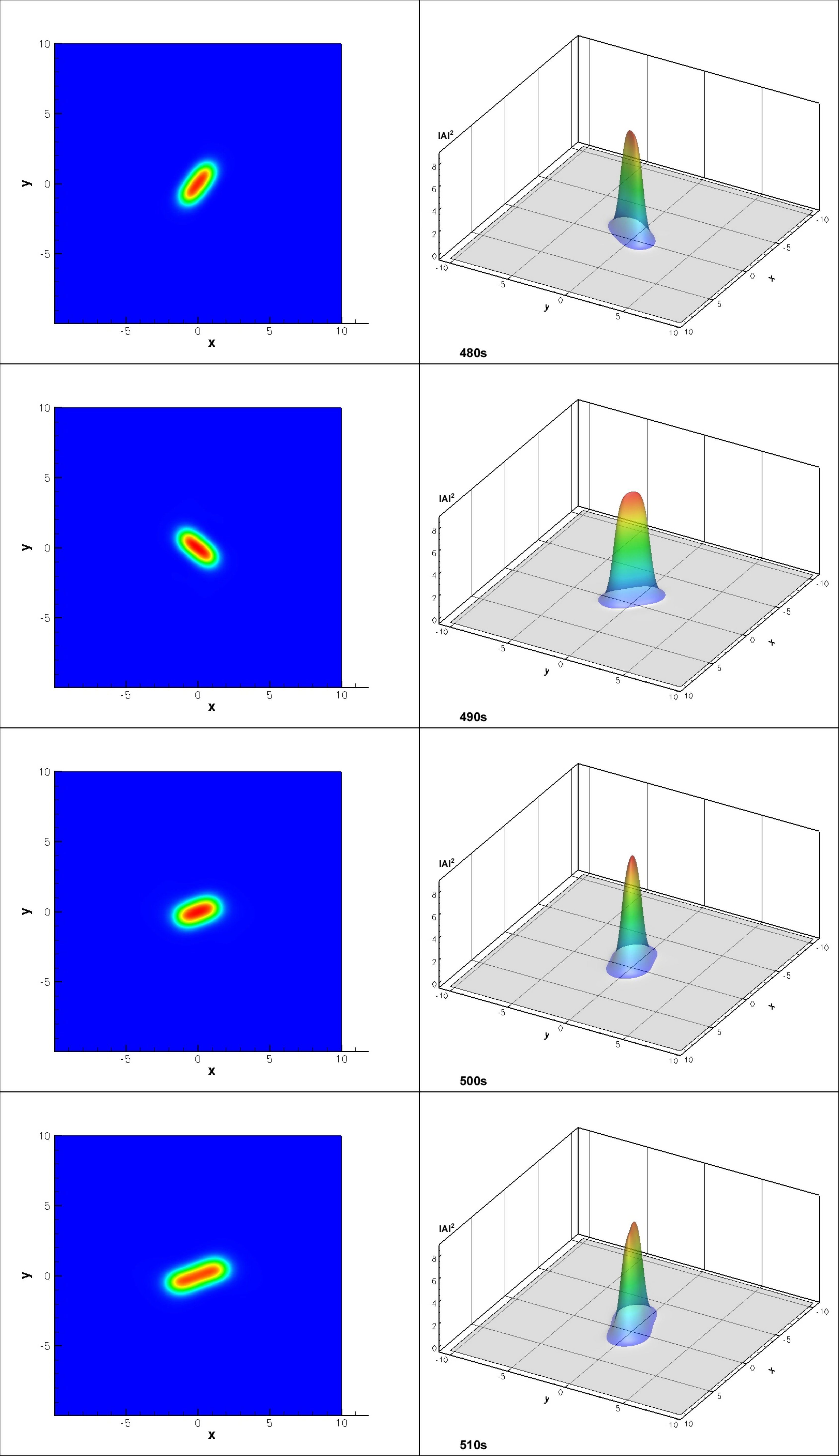}
\caption{During the pulsating state $t \in [480s,520s]$ the soliton elongates alternatively in different directions. Parameters: row 2 of Table \ref{TabParam}.}
\label{Pulse2}
\end{figure}
For $b_3=0.37$ let us analyze in detail the behavior of the  soliton in the pulsating state of $t \in [480 \,s, 500 \,s]$, see  Fig. \ref{Pulse2}. For the first $400 \,s$, the evolution of the solution looks similar to the stationary soliton, see Fig. \ref{Normal}. While the soliton is stabilized at $t=430s$,  a bifurcation appears to make the structure evolve toward a new higher state: the energy $Q$ of the soliton is now pulsating. The structure still has radially symmetric profile when $Q$ takes its maximum, but elongates alternatively in different directions when $Q$ takes its minimum value. Indeed, this state is not stable for this set of parameters and evolves into another stable soliton which keeps the elongated shape, as if it was a double bell-shaped soliton. The energy still pulsates but with a very little amplitude and the soliton rotates  around its center, see Fig. \ref{Pulse2}.

\subsubsection{Ranges of parameters and bifurcations}
\begin{figure}
\centerline{\includegraphics[width=6cm]{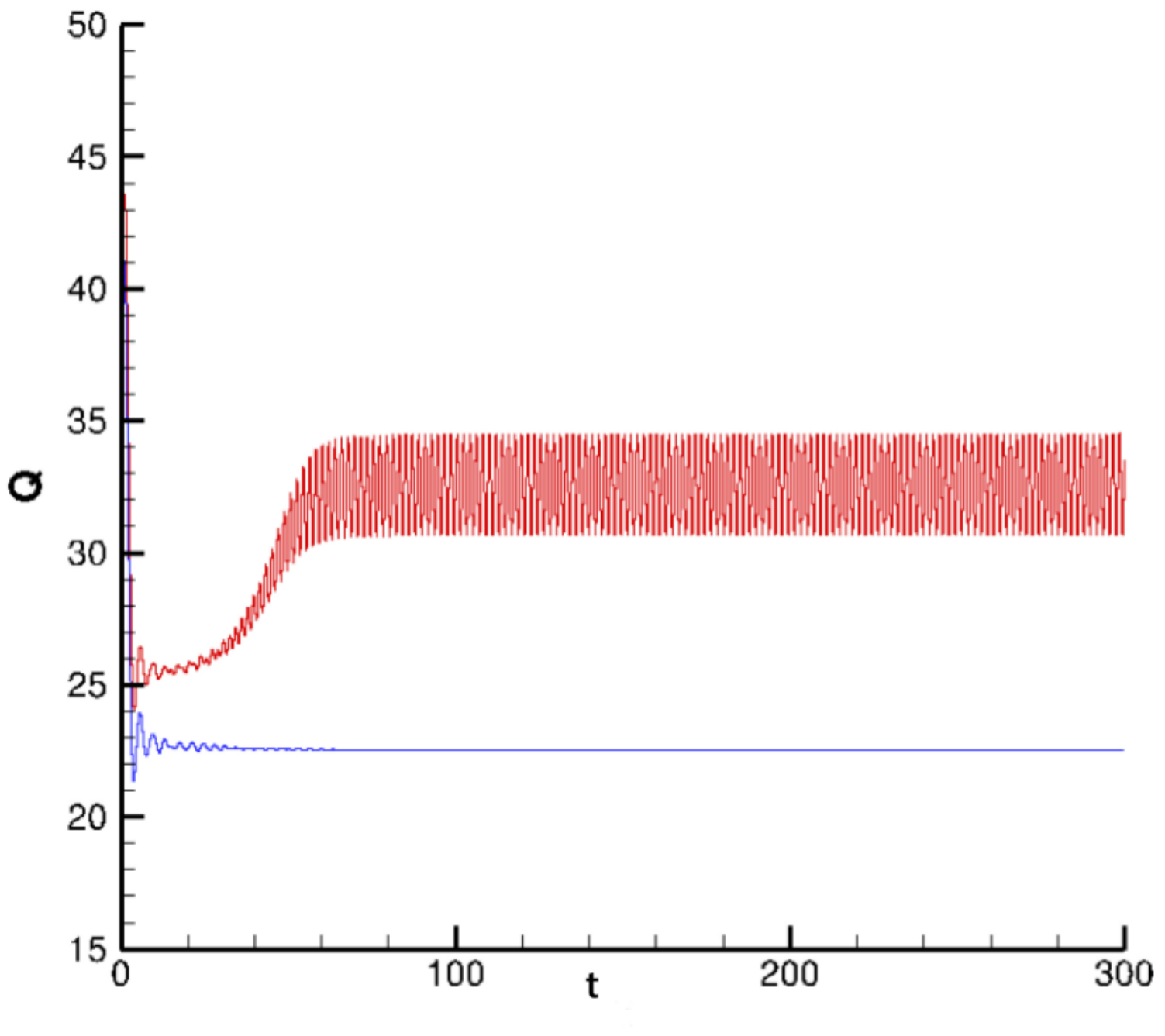}}
\caption{Bifurcations for pulsating soliton for  $b_3=-0.36$ (top energy in red) that becomes  stationary soliton when $b_3=-0.35$ (bottom energy in blue).}
\label{puls4}
\end{figure}

For the ranges of parameters explored, the pulsating soliton experiences another type of bifurcation that leads to a stationary soliton as we change $b_5$. This parameter seems to be  very sensitive  for the pulsating soliton because for small changes the top pulsating energy becomes the stationary lower energy, and hence the pulsating becomes stationary soliton, see  Fig. \ref{puls4}.   Indeed, whereas for $b_5 \in [-0.05, -0.02]$ the soliton is pulsating, in  $[0.06, 0.18]$ soliton is stationary.  Between these two intervals, in $[-0.02, 0.04]$  solitons are unstable. Table \ref{TabPuls} gives an overview of the parameter ranges explored for the pulsating solitons.
\begin{table}
\begin{center}
\begin{tabular}{|c|c|}
\hline Parameters & Pulsating  \\
\hline
$b_1$ & [0.015, 0.035]\\
\hline
$\epsilon$ & [-0.095, -0.045]   \\
\hline
$b_3$ & [-0.4, -0.36]  \\
\hline
$b_5$ & [-0.05, -0.02]  \\
\hline
$c_5$ & [-0.09, -0.08] \\
\hline
\end{tabular}
\end{center}
\caption{Parameter ranges of existence of pulsating  solitons.}
\label{TabPuls}
\end{table}

%%%%%%%%%%%%%%%%%%%%%%%%%%%%%%%%%%%%%%%%%%%%%%%%CCCCCCCCCCCCCCCCCCCCCCCCCC

%%%%%%%%%%%%%%%%%%%%%%%%%%%%%%%%%%%%%%%%%%%%%%%%%%
\subsection{Circular and Elliptic  vortex (spinning)}
\subsubsection{Description}
\begin{figure}[htbp]
\centering
\includegraphics[width=8cm]{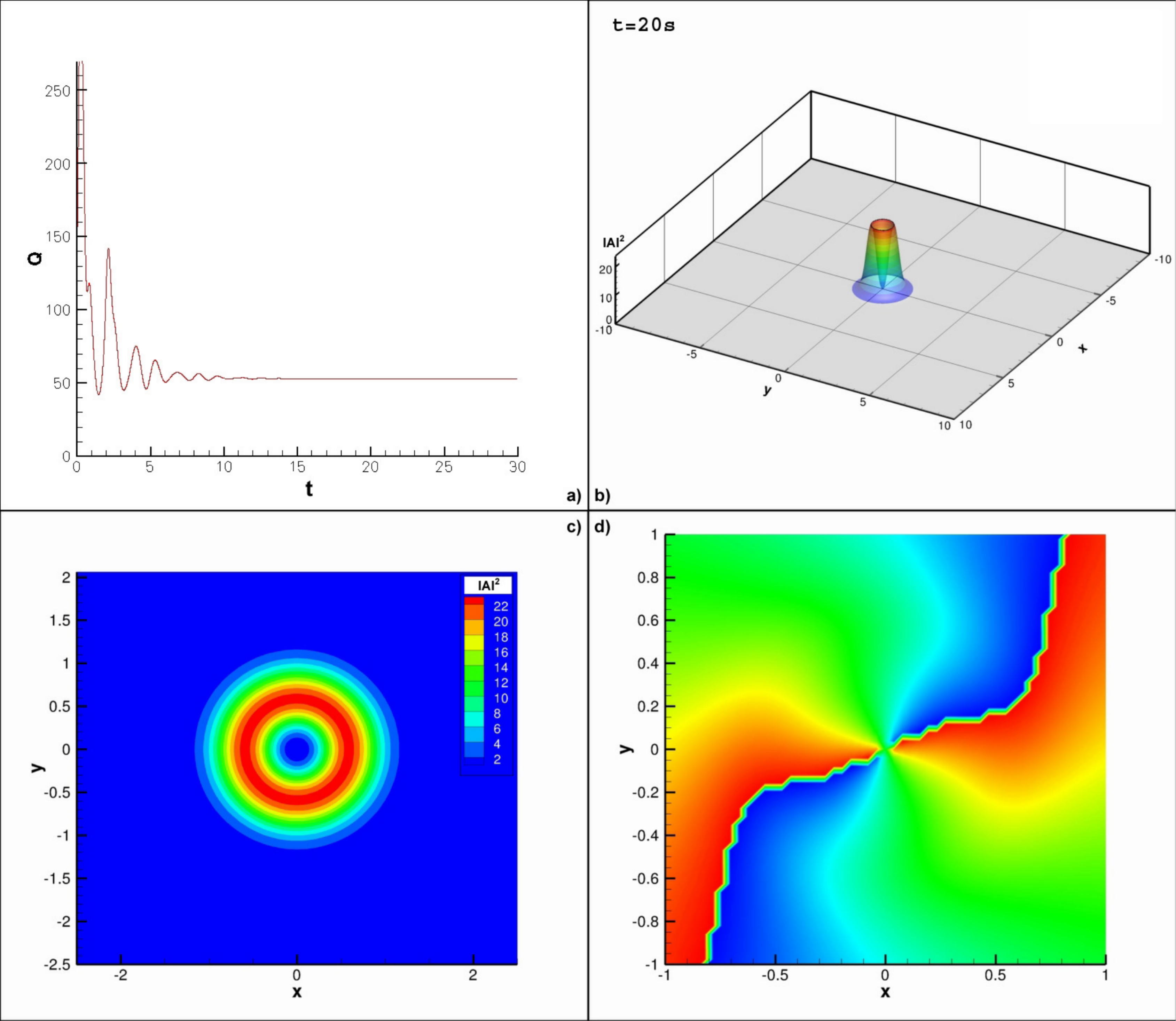}
\caption{Circular ring vortex soliton with radially symmetry. Top left: Energy. Top right: Ring vortex soliton at $t=20s$. Bottom left: Contour plot of $|A|^2$. Bottom right: Phase plot that indicates rotation at $t=20 s$.  Parameters: row 3 of Table \ref{TabParam}.}
\label{Ring}
\end{figure}

Stable ring vortices were found before numerically  by \cite{Skarka:1,Crasovan}. Using a variational formulation similar to \cite{Mancas:4} an analysis of 2D doughnut-shaped pulses with phase in the form of a rotating spiral have been found numerically also by Mihalache {\it et. al.} \cite{Crasovan:2}. For this class we  choose a circular ring type structure with vorticity $m=1$, amplitude $A_0=2.5$, widths $\sigma_x=\sigma_y=1$, and parameters  from row 3 of Table \ref{TabParam}. First, we found a radially symmetric ring vortex soliton shown in Fig. \ref{Ring} top two panels. Even though the structure oscillates before converging to a stable value of the energy as it is with the stationary soliton,  since now we have  a phase the soliton is spinning. The rotation has a fixed angular velocity and can occur in either direction clockwise or counterclockwise, which is determined by the initial condition, see Fig. \ref{Ring} bottom right panel.

\begin{figure}[htbp]
\centering
\includegraphics[width=8cm]{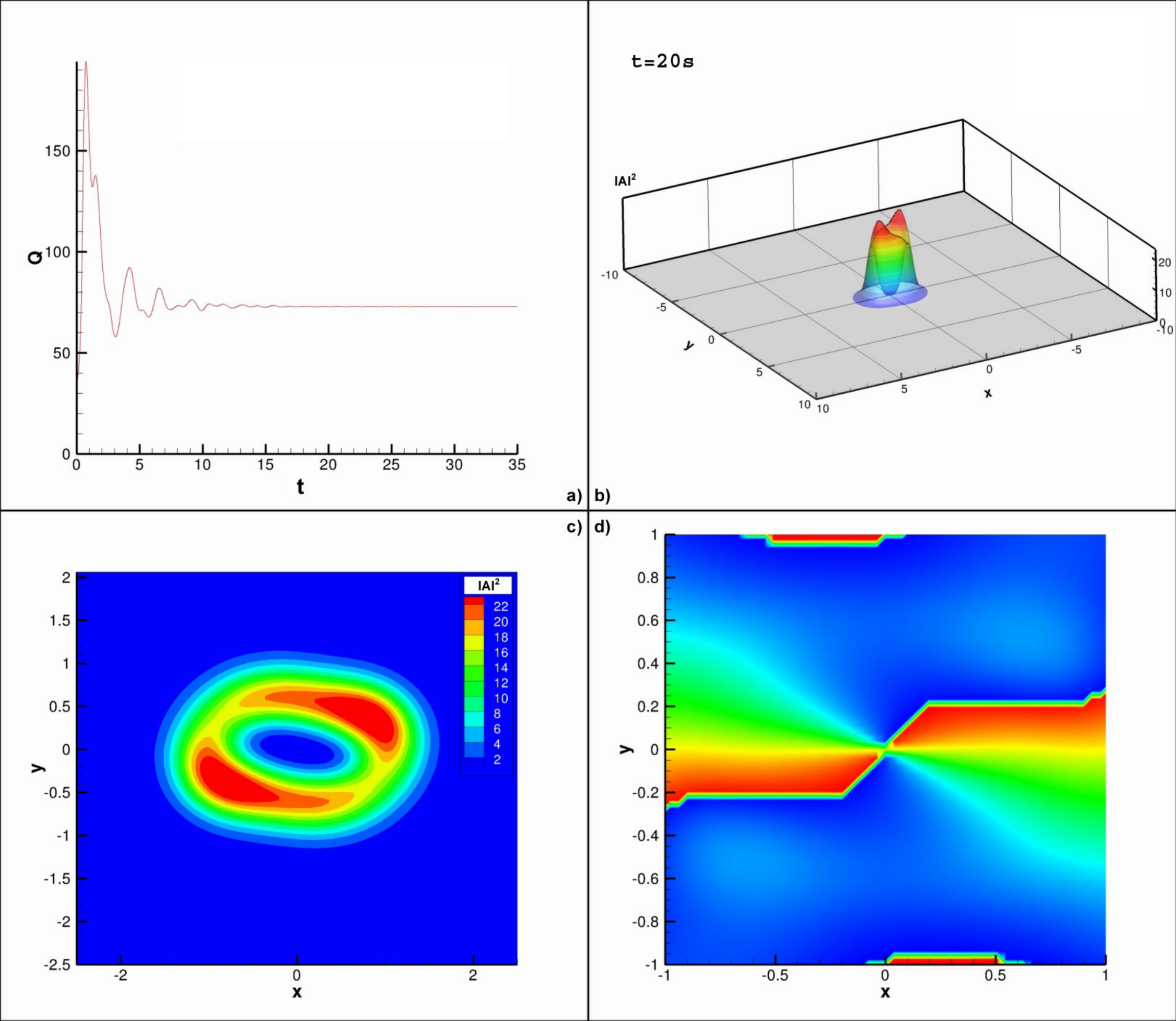}
\caption{Elliptic ring vortex soliton (no radial symmetry). Two peaks appear at the top of the structure, while the phase keeps spinning.  Parameters: row 3  of Table \ref{TabParam}.}
\label{RingEllip}
\end{figure}

Next, we introduced for the same set of parameters,  an elliptic shape initial condition, which breaks the  symmetry, by choosing $\sigma_x=0.15$, $\sigma_y=0.85$. This new  condition leads to another kind of soliton with an elongated shape and two peaks diametrically opposed. This soliton which lacks radial symmetry has two peaks of amplitude diametrically opposed at the top of the structure \cite{Akhmediev:2}. The energy first oscillates before converging to a fixed value and remains stationary, see Fig. \ref{RingEllip} top panels, while from the bottom right panel we deduce that the soliton is still spinning, but not as fast as the circular one. 

\begin{figure}[htbp]
\centering
\includegraphics[width=9cm]{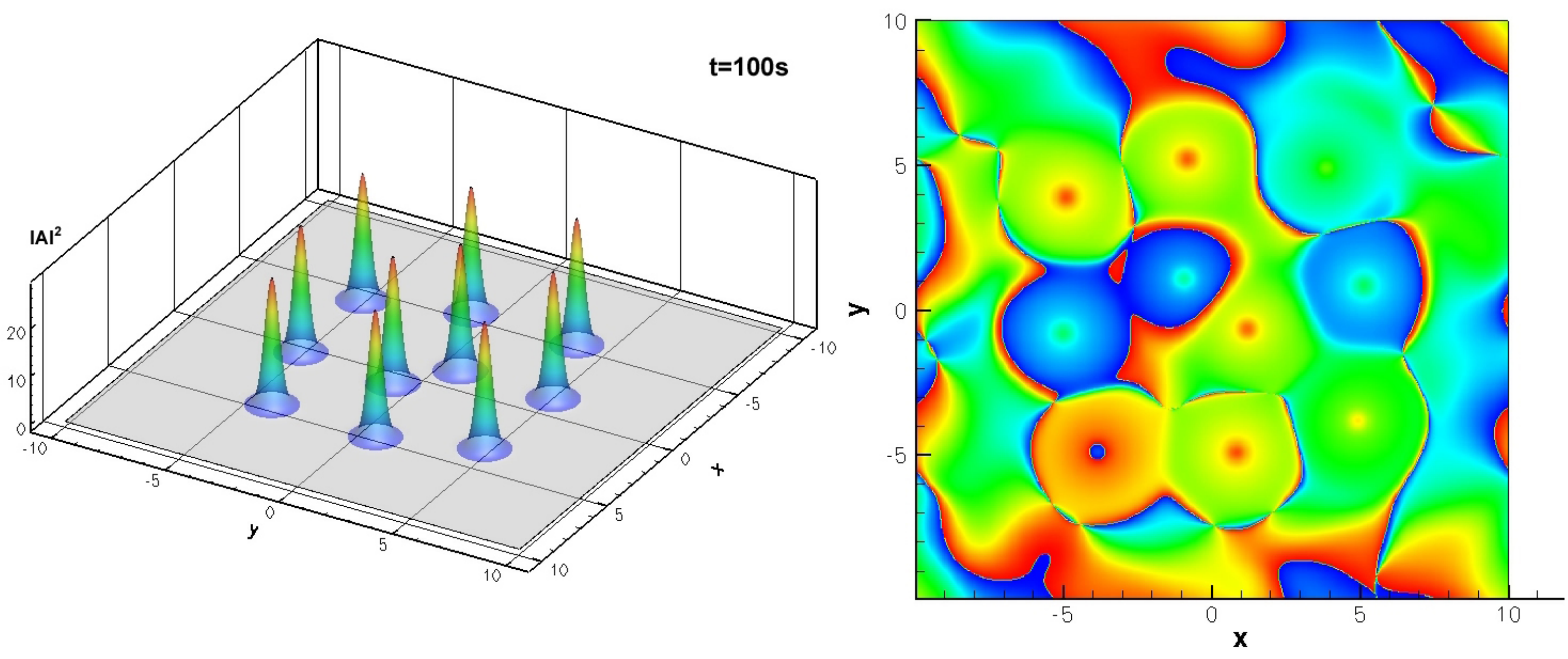}
\caption{Unstable vortex circular ring. Left: 10 bell-shaped solitons appeared due to the destruction of the initial ring soliton. Right:  the phases of the solitons are not spinning. The parameters of the initial shape are: $A_0=3.0$, $\sigma_x=\sigma_y=0.15$. Parameters: row 3 of Table \ref{TabParam}.}
\label{filament}
\end{figure}

\begin{figure}
\centerline{\includegraphics[width=8cm]{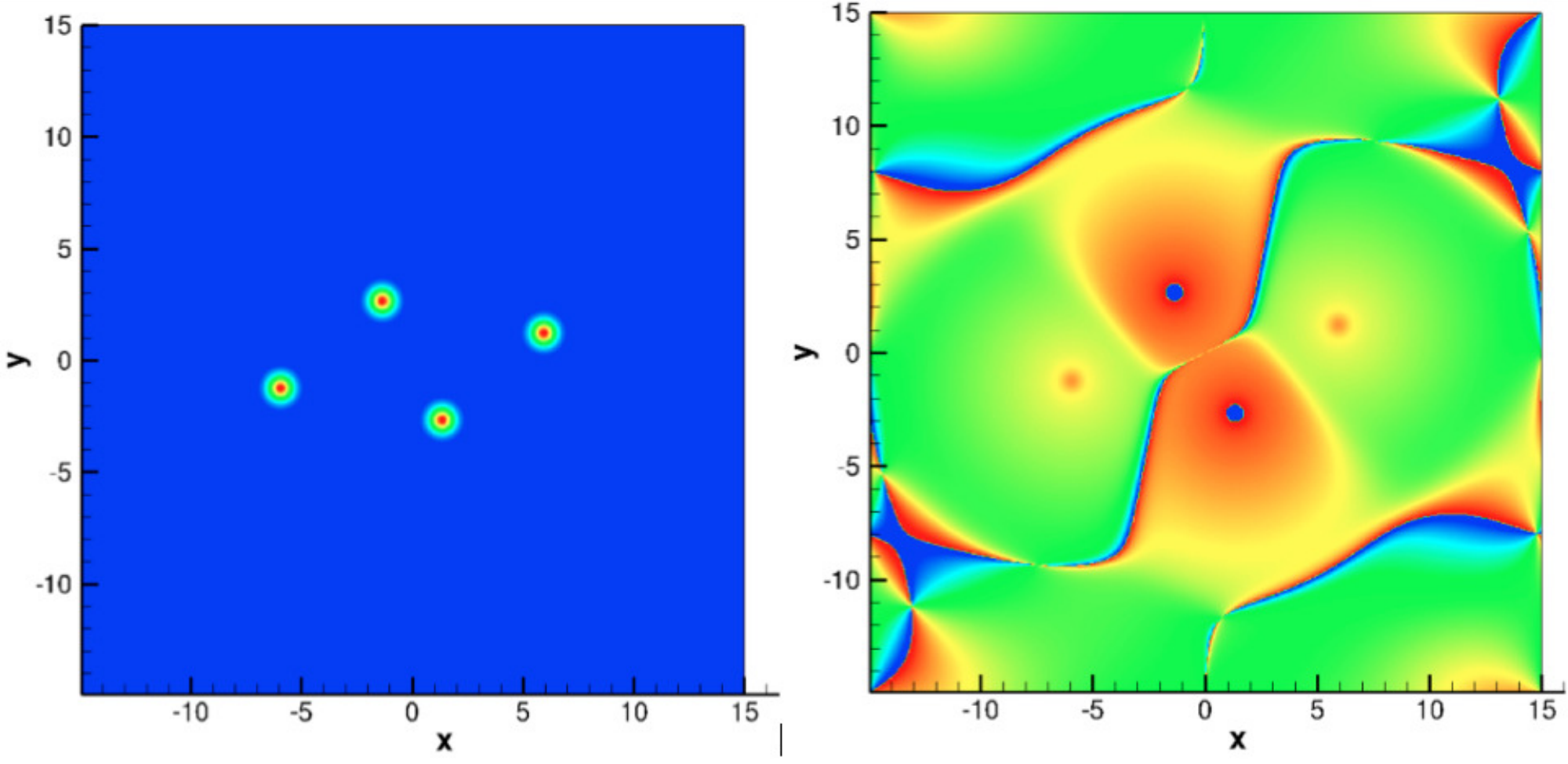}}
\centerline{\includegraphics[width=8cm]{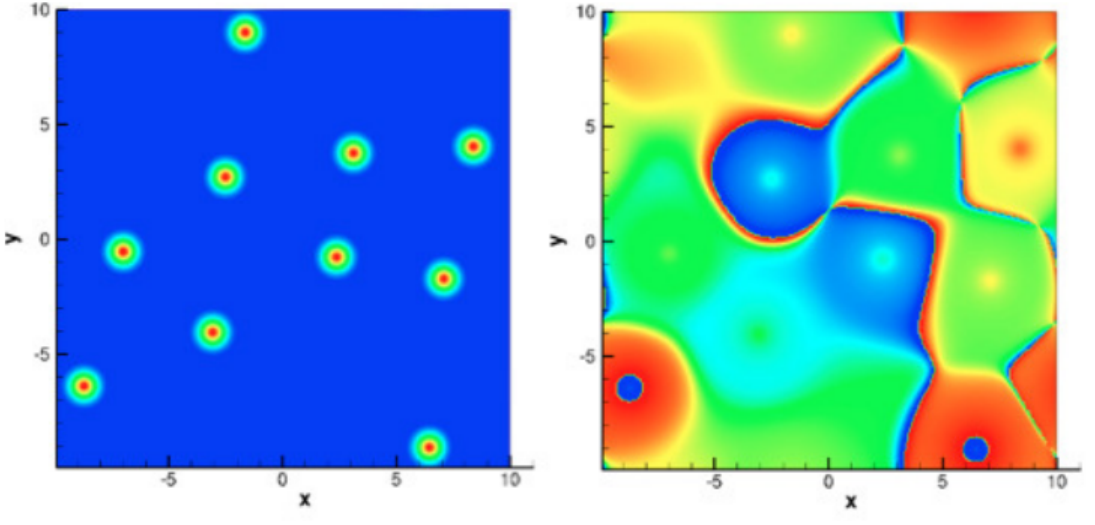}}
\caption{Filament soliton with elliptic vortex initial condition $m=1$, $m=2$, $b_5=0.18$, $t=100s$. Left: contour plot of $|A|^2$. Right: phase plot that shows no spin.}
\label{Filab}
\end{figure}

Ordinary bell-shaped solitons with zero-vorticity can coexist for the same parameters, but with lower energy. Stable ring vortices, filamentation and also the unstable ones  have also been found previously by \cite{Skarka:1,Nody2} using  using a variational formulation. Increasing the amplitude and choosing the initial condition to be circular, $A_0=3.0$, $\sigma_x=\sigma_y=0.15$ results in a splitting of the initial beam in several bell-shaped solitons which are non spinning. If a ring vortex soliton loses its stability, it will be transformed into several bell-shaped solitons via multiple bifurcations. That is what we have found, by chance, while looking for the spinning ring vortex soliton, see Fig. \ref{filament}.  As it has been shown in \cite{Akhmediev:2}, this class may be unstable and leads to chaos. We also noticed that for the same set of parameters, the beams stop splitting and the soliton appears to be stable. We obtained both filament solitons for vorticity $m=1$, and $m=2$, see  Fig. \ref{Filab}. For other results with higher vorticities, see \cite{Ankiewicz:2}.

\subsubsection{Ranges of parameters and bifurcations}
\begin{figure}
\centerline{\includegraphics[width=8cm]{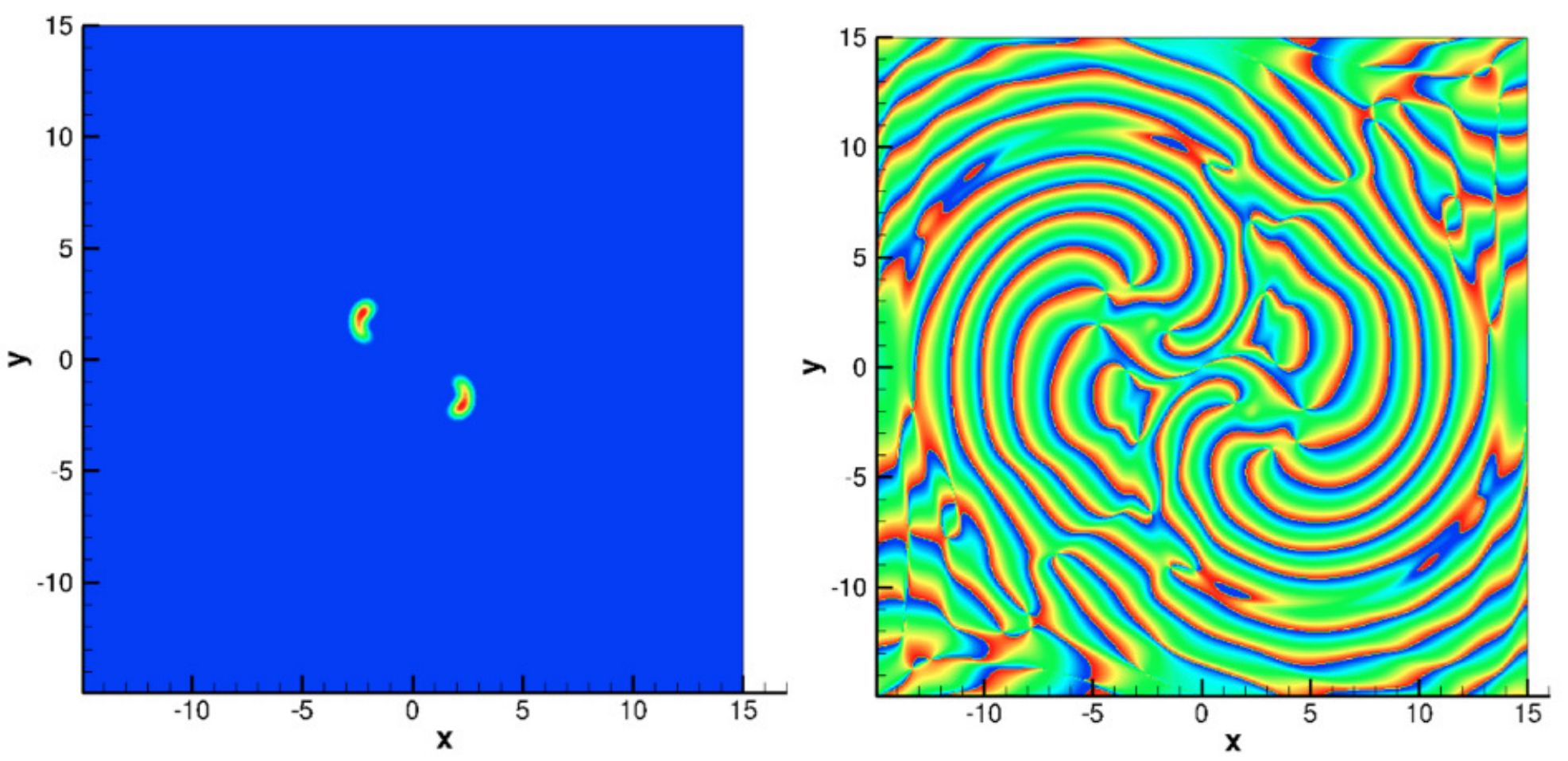}}
\caption{Spinning ``bean-shaped'' soliton with elliptic vortex initial condition $m=1$, $\epsilon=-0.17$, $t=50s$. Left: contour plot of $|A|^2$. Right: phase plot.}
\label{Filac}
\end{figure}

Elliptic spinning and filament solitons undergo bifurcations which result in stable solitons over a large parameter range around the initial sets of parameters. By varying one parameter from row 3 of Table \ref{TabParam} several bifurcations happen and other stable solitons appear with an energy evolution similar to these presented in Figs. \ref{Ring}, \ref{RingEllip}. The main bifurcations observed lead to the transformation of the initial beam in a stable spinning ring, filament or quite exotic spinning ``bean-shaped'' solitons, see Fig. \ref{Filac}. As far as ranges of parameters, we were only able to determine the ranges of the stable spinning solitons, and the results are shown in Table \ref{TabSpin}.

\begin{table}
\begin{center}
\begin{tabular}{|c|c|}
\hline Parameters & Spinning  \\
\hline
$b_1$ & [0.035, 0.235]  \\
\hline
$\epsilon$ & [-0.3, -0.1]  \\
\hline
$b_3$ & [-0.9, -0.7] \\
\hline
$b_5$ & [-0.02, 0.18] \\
\hline
$c_5$ & [-0.22, -0.02] \\
\hline
\end{tabular}
\end{center}
\caption{Parameter ranges of spinning stable solitons.}
\label{TabSpin}
\end{table}

%%%%%%%%%%%%%%%%%%%%%%%%%%%%%%%%%%%%%%%%%%%%%%%%%%%%%%%%

%%%%%%%%%%%%%%%%%%%%%%%%%%%%%%%%%%%%%%%%%%%%%%%%%%%%%%%%%

%%%%%%%%%%%%%%%%%%%%%EEEEEEEEEEEEEEEEEEEE11111111111111111111111111111111111111

\subsection{Exploding solitons}
\subsubsection{Description}
\begin{figure}[htbp]
\centering
\centerline{\includegraphics[width=8cm]{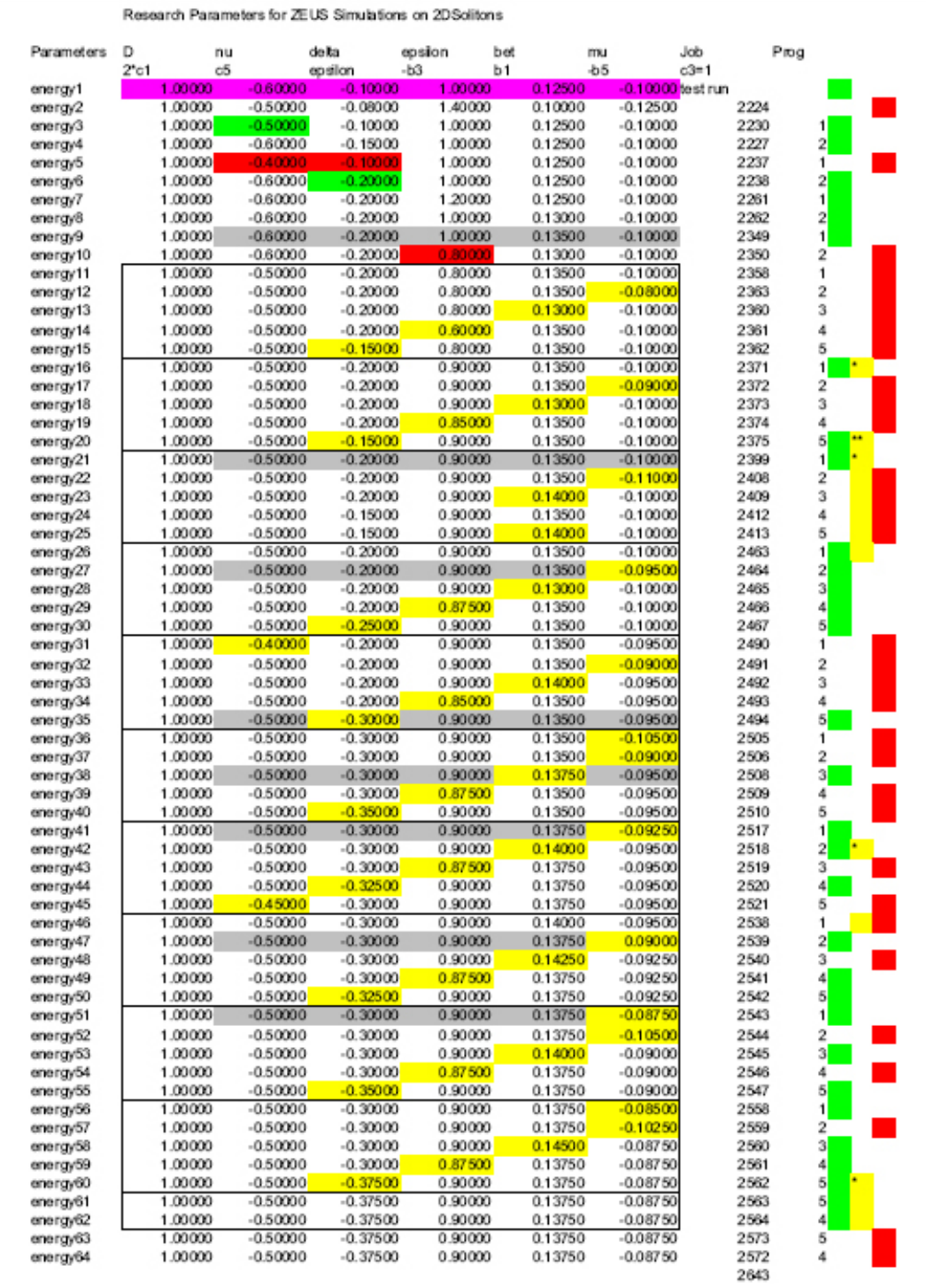}}
\caption{Parameters used to simulate the exploding solitons} \label{Figure2}
\end{figure}

\begin{figure}[htbp]
\centering
\centerline{\includegraphics[width=6cm]{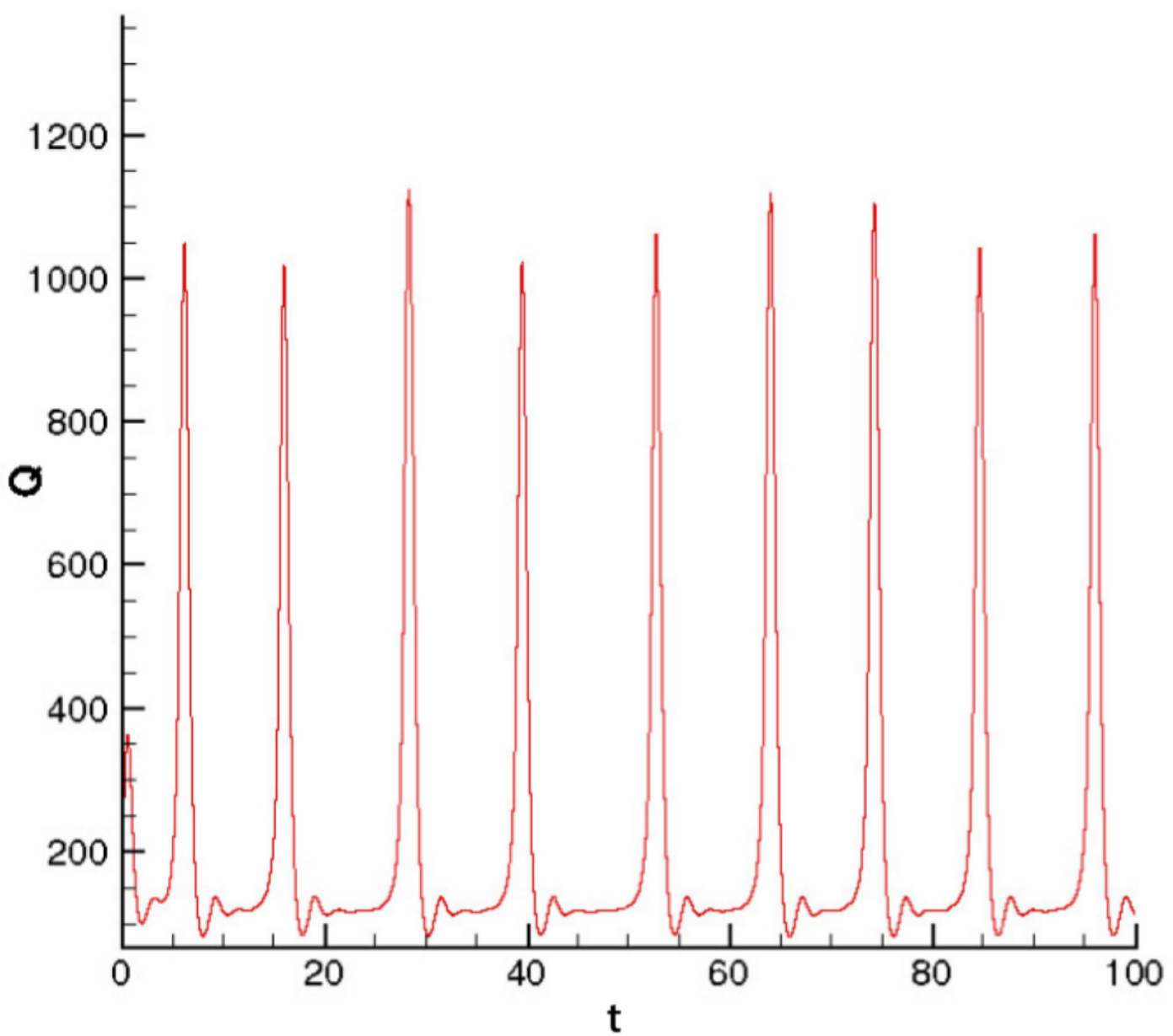}}
\caption{Exploding soliton: energy is periodic with high bursts almost every $12s$.  Parameters: row 4 of Table \ref{TabParam}.}
\label{ExplE}
\end{figure}

\begin{figure}[htbp]
\centering
\centerline{\includegraphics[width=8cm]{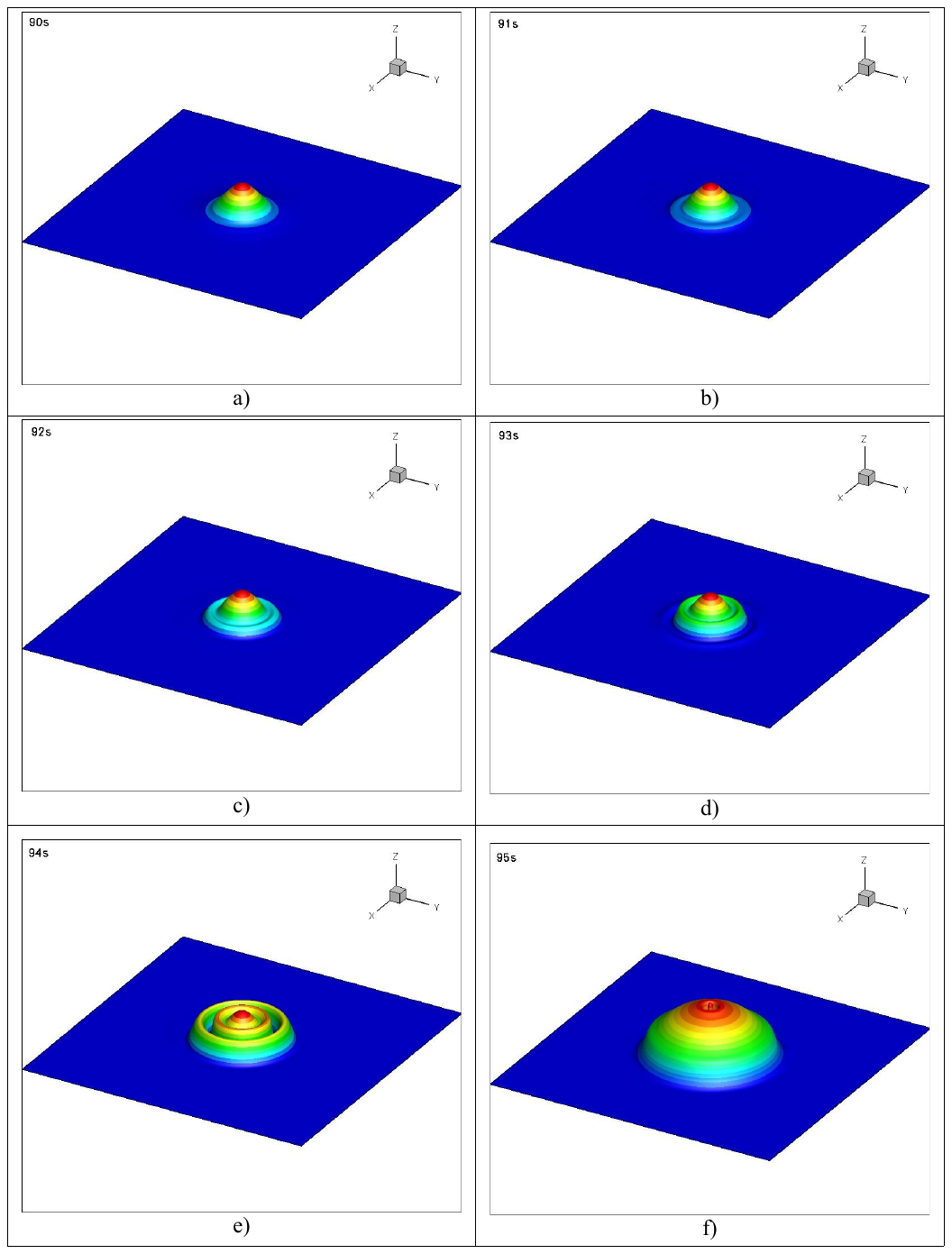}}
\caption{Exploding soliton at $t=90\,s$, the shape is smooth. Then, circular waves appear and grow. Finally, soliton explodes around $t=94 \, s$ and collapses after which it restores back to its initial shape.  Parameters: row 4 of Table \ref{TabParam}.}
\label{ExplA}
\end{figure}

This class was first mentioned in \cite{Akhmediev:1,Akhmediev:2,Soto:2,Soto:4,str}, and later by others \cite{Crasovan:3,Ora:1,Ora:2,Ora:3}, and \cite{Ora:4} for symmetric and asymmetric transients.   For this class we include a sample of how the simulations sets look like.  We initially computed  $64$ simulations and all changes in parameters were recorded in a chart, see Fig. \ref{Figure2}, within the $7$ dimensional space given by $c_1 = .5$, $c_5 \in [ -0.6, -0.4]$, $\epsilon \in [-0.08, -.375]$, $b_3 \in [-0.06, -1.4]$, $b_1 \in [ 0.1, 0.145]$, $b_5 \in [0.08, 0.125]$, $c_3 = 1$.
For each set, the energy $Q(t)$ was monitored. To obtain a correct exploding soliton, we need peaks of energy that exceed the average value by a large margin. Also, this energy must be almost periodic, since after each explosion the soliton must recover its original shape.  See the energy $Q(t)$ in Fig. \ref{ExplE} computed using the parameters of row 4 of Table \ref{TabParam}. The evolution starts from a circular  initial condition of Gaussian profile with amplitude $A_0=3$, and widths $\sigma_x=\sigma_y=0.3$.
The main feature of this soliton is that as it propagates  its slopes become covered with small scale instabilities which seem to move downwards. The soliton explodes intermittently,
thus resulting in significant bursts of power above the average,
recovering the initial radially cylindrical shape after
each explosion \cite{Ankiewicz:2, Crespo:1}, see Fig. \ref{ExplA}. As it was shown for the 1D exploding solitons \cite{Ora:2,Ora:3},
this unusual dynamics appears as a result of an instability.
As we move further from this boundary, the explosions become
more violent, and, as a result more than one beam can
be generated in some cases. This gives rise to  very complicated
dynamics, ending up with the whole numerical grid
filled with the solution in transit to chaos. 

At first, the shape seems to be stable with constant energy and smooth bell-shaped cross section at $t= 90\,s$. Often, circular waves ripple  from the center to the exterior and vanish as if energy were dissipated in the boundaries of the solitons, see $t=92 \,s$. Suddenly, and periodically, the soliton ``explodes", see $t=94 \, s$: its shape grows, so do the energy, until it collapses and goes back to the original shape, see  $t=90 \,s$. This completely chaotic, but well-localized structure restores to a perfect solitons shape after the main burst.  The periodic evolution of the energy, with high bursts of energy can be seen in Fig. \ref{ExplE}, while the exploding soliton as it progresses can be seen in Fig. \ref{ExplA}. This process never repeats itself in successive periods, however, it always returns to the same shape. Hence, the essential features of the explosions, both observed numerically in 1D  by \cite{Crespo:1,Akhmediev:1} are: explosions occur intermittently, the explosions have similar features but are not identical, and explosions happen spontaneously triggered by perturbations.

%%%%%%%%%%%%%%%%%%%%%%%%%%%%%FFFFFFFFFFFFFFFFFFFFFFFFFF%%%%%%%%%%%%%%%%%%%%%%%%%%%%%%%%%%

\subsubsection{Ranges of parameters and bifurcations}
For determining the ranges of the exploding solitons, we encountered two cases: the stable exploding solitons, and unstable ones for which the energy diverges to infinity. The later type never recovers their shape, and appears probably when the features of the system and the medium (defined by the parameters of the CCQGLE) cannot compensate the energy bursts, and hence let the energy grow without bound.

\begin{figure}
\centerline{\includegraphics[width=6cm]{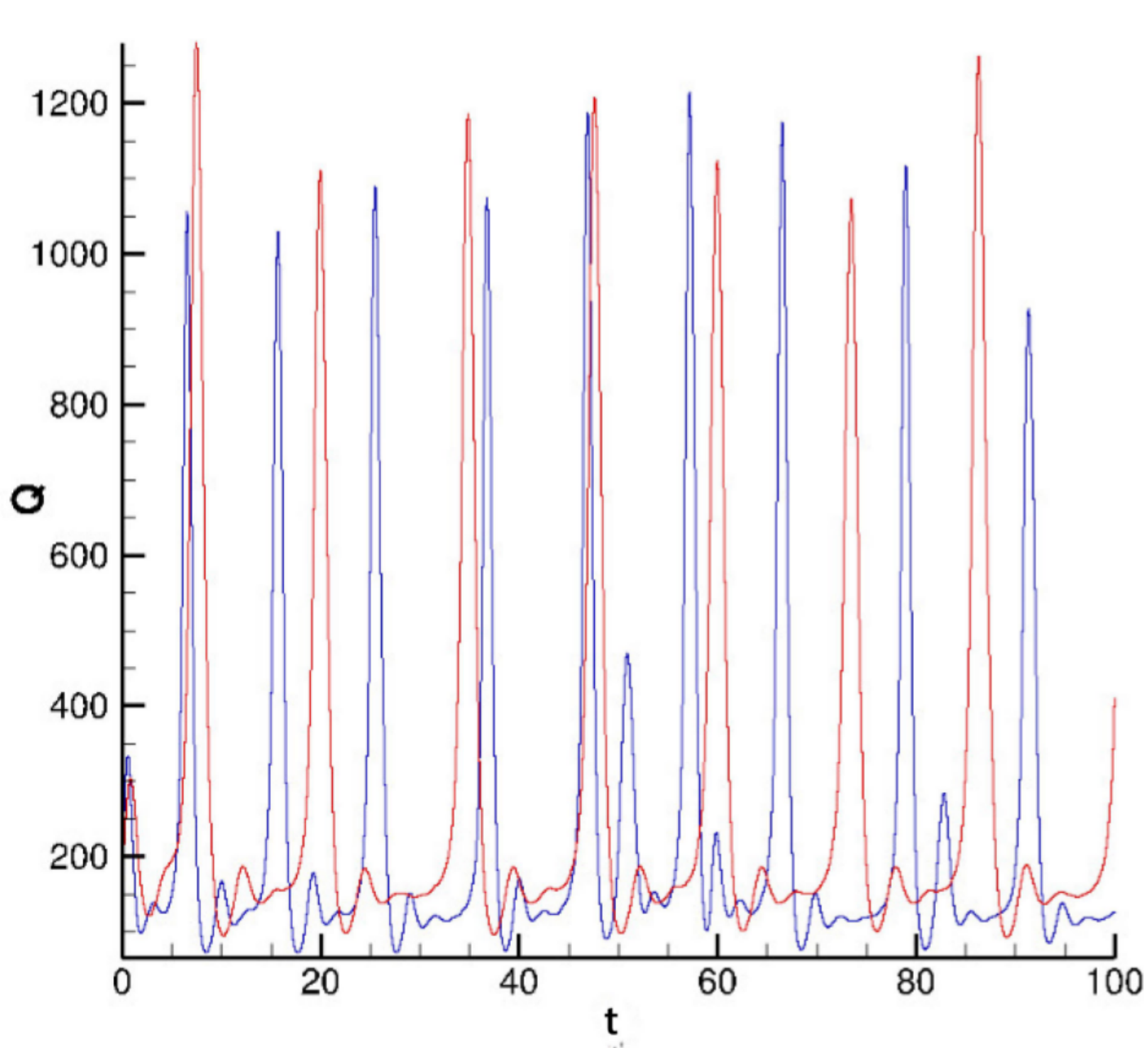}}
\caption{Energies of two exploding solitons obtained with $b_3=-0.96$ (blue) and $b_3=-0.81$ (red): the explosions are shifted with different bursts.}
\label{expl3}
\end{figure}

\begin{figure}
\centerline{\includegraphics[width=7cm]{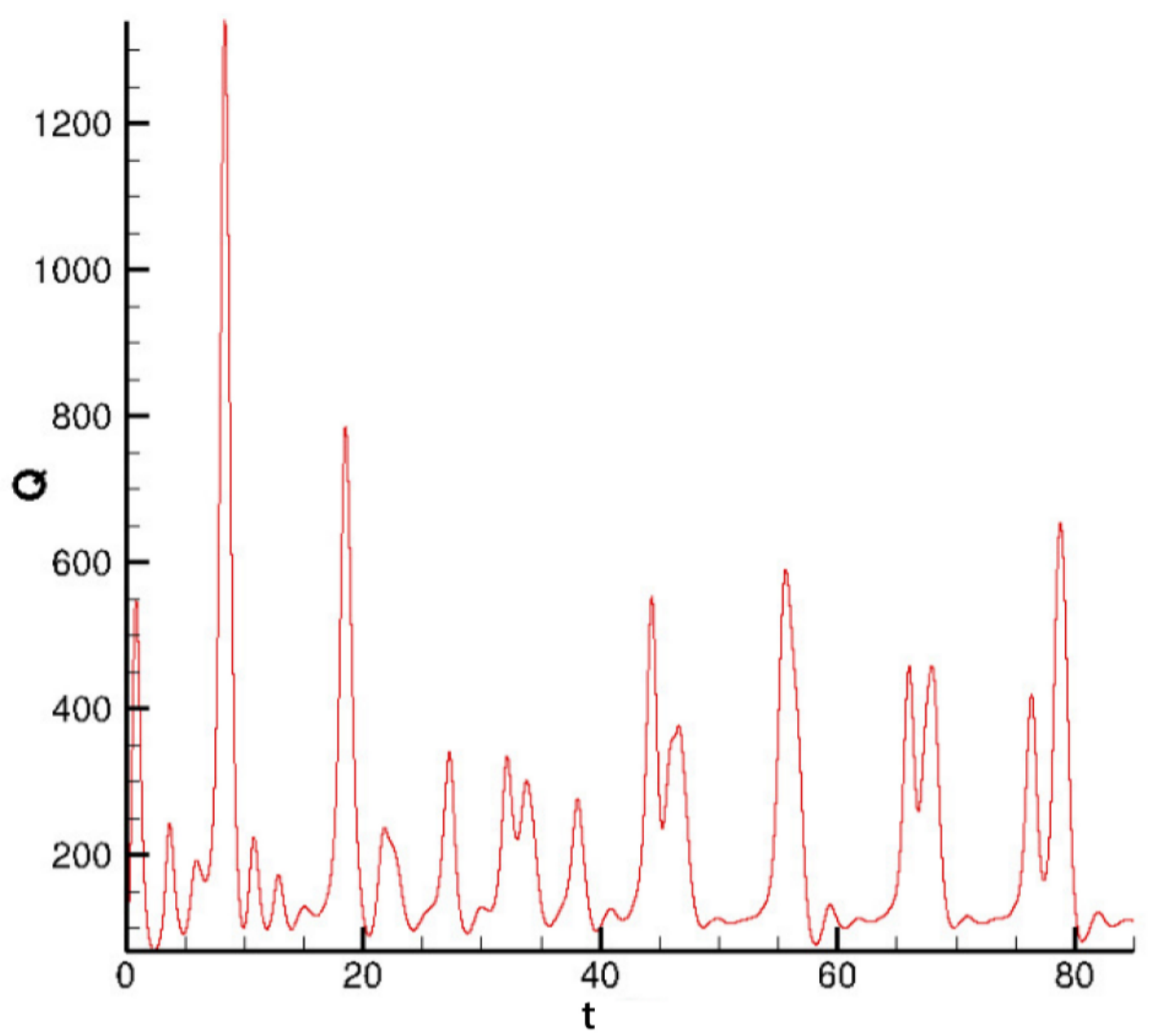}}
\caption{Disorganized exploding soliton starting as a ring vortex with $m=1$, and $b_5=0.14$.}
\label{expl4}
\end{figure}

As it was explained previously, small changes in some parameters dramatically change  features of the exploding soliton. As it is shown in Fig. \ref{expl3}, the effect of changing $b_3$ shows an energy evolution with random amplitudes and shifts. Also, depending on the initial conditions, the explosions may be organized or disorganized, as we can see in Fig. \ref{expl4}. The disorganized explosions are due to  the  asymmetric structure that evolves to an asymmetric exploding soliton \cite{Ora:1,Ora:4}.

A ring vortex initial condition has been tested in the same ranges of parameters. Contrary to pulsating solitons which did not exist with this initial shape, exploding solitons seemed to be existing. However, even when explosions and bursts happened, the behavior of such a soliton was extremely unpredictable. In fact, the evolution of the energy is very random, explosions do not appear periodically and the amplitudes of these explosions are relatively different, see Fig. \ref{expl4}. We assume that this initial condition leads to an exploding soliton that is on transition to chaos. The ranges of  existence for organized explosions are shown in Table \ref{TabExp}, while for disorganized are shown in Table \ref{TabExp2}.

\begin{table}
\begin{center}
\begin{tabular}{|c|c|}
\hline Parameters & Range \\
\hline
$b_1$ & [0.135, 0.235] \\
\hline
$\epsilon$ & [-0.3, -0.1]  \\
\hline
$b_3$ & [-1.0, 0.8]   \\
\hline
$b_5$ & [0.09, 0.15]  \\
\hline
$c_5$ & [-0.6, -0.4]  \\
\hline
\end{tabular}
\end{center}
\caption{Parameter ranges of existence for organized exploding solitons.}
\label{TabExp}
\end{table}

\begin{table}
\begin{center}
\begin{tabular}{|c|c|}
\hline Parameters & Range  \\
\hline
$b_1$ & [0.085, 0.135] \\
\hline
$\epsilon$ & [-0.26, -0.1]  \\
\hline
$b_3$ & [-1.0, 0.8]   \\
\hline
$b_5$ & [0.13, 0.15]  \\
\hline
$c_5$ & [-0.6, -0.5]  \\
\hline
\end{tabular}
\end{center}
\caption{Parameter ranges of existence for disorganized exploding solitons}
\label{TabExp2}
\end{table}

%%%%%%%%%%%%%%%%%%%%%FFFFFFFFFFFFFFFFFFFF%%%%%%%%%%%%%%%%%%%%%%%%%%%%%%%%%%%

\subsection{Creeping solitons}
\subsubsection{Description}
\begin{figure}
\centerline{\includegraphics[width=8cm]{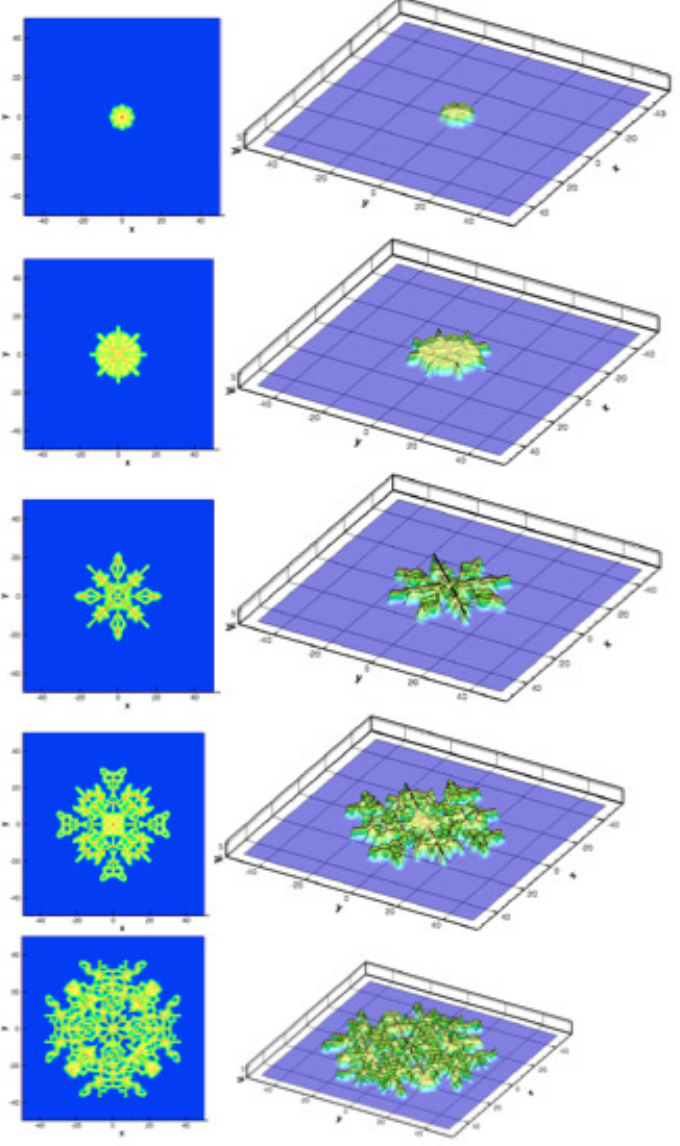}}
\caption{Creeping soliton: Initial Gaussian beam is spreading like a fractal from top to bottom.}
\label{cre}
\end{figure}

Creeping soliton dynamic behavior also starts from a Gaussian initial condition with radial symmetry, with amplitude $A_0=3.0$,  widths $\sigma_x=\sigma_y=0.3$, and initial paramteres from row 5 of Table \ref{TabParam}.   In 2D, creeping solitons behave like 2D tumors while in 1D creeping solitons move to only one direction with constant velocity, see \cite{5,Crespo:1}. In fact, they spread in all directions until filling the domain, even though their evolution is organized. Thus, the energy of a 2D creeping soliton is very often characterized by a constant raise, hence precautions must be taken when the size of the domain is defined. Before spreading and filling the numerical grid, creeping solitons present a very complex dynamic evolution pattern by taking surprising geometrical shapes that resemble fractals, see Fig. \ref{cre}. They are still chaotic but localized in the domain and tend to show  symmetry. In the 1D case, it has been proved \cite{5} that the region of parameters where creeping solitons exist is filled which a rich variety of bifurcations between, stationary, pulsating and creeping solitons.

\subsubsection{Ranges of parameters}

Two main bifurcations were observed by varying  parameters of row 5 of Table \ref{TabParam}.  By using the Gaussian beam as initial condition, the first type of creeping takes the shape of a ring which is expanding very fast (when the energy is constantly increasing) and leads to chaos see Fig. \ref{cre}, while the second one (when the energy converges to a finite value)  leads to stationary solitons as in Section 4.  The parameters ranges are presented in Table \ref{TabCreep}.

\begin{figure}
\centerline{\includegraphics[width=8cm]{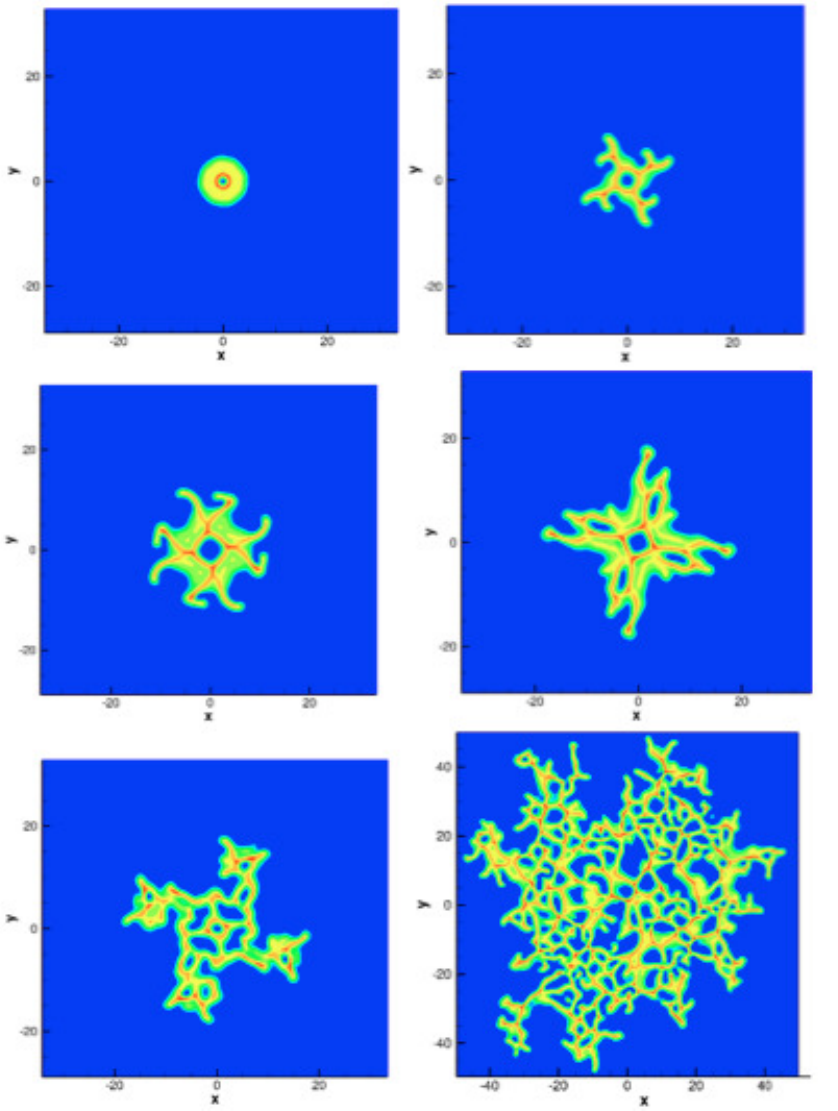}}
\caption{Evolution of the shape of a propeller soliton.}
\label{prop}
\end{figure}

\begin{figure}
\centerline{\includegraphics[width=7cm]{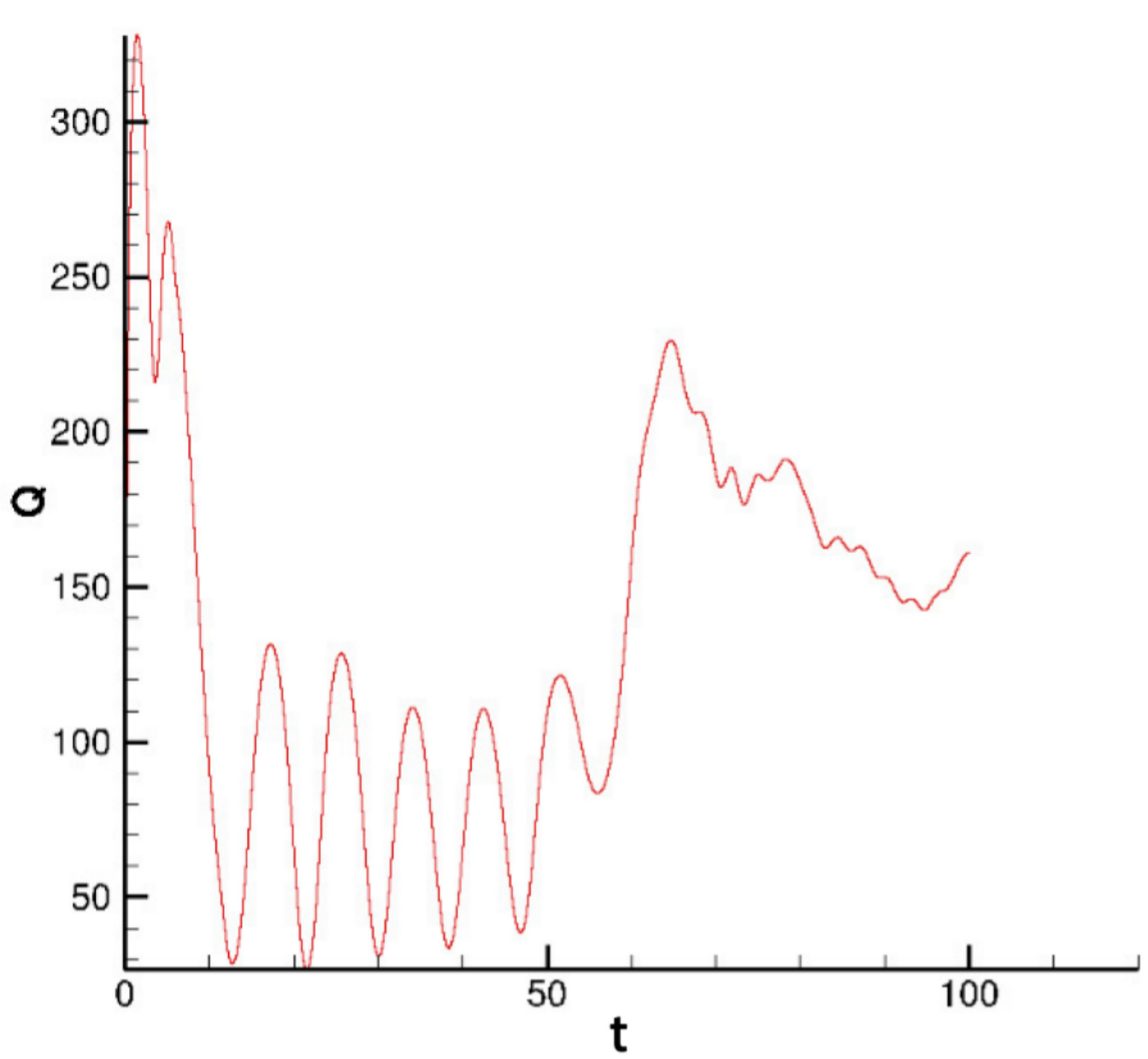}}
\caption{Evolution of the energy of a spreading spinning soliton next to the bifurcation boundary.}
\label{prop2}
\end{figure}

\begin{table}
\begin{center}
\begin{tabular}{|c|cc|}
\hline Parameters & Creeping  & Stationary \\
\hline
$b_1$ & [0.08, 0.14] & not found \\
\hline
$\epsilon$ & [-0.13, -0.09] & [-0.3; -0.14]  \\
\hline
$b_3$ & [-1.3, -1.29] & [-1.28; -1.11]  \\
\hline
$b_5$ & [0.3, 0.31] & [0.32, 0.36]  \\
\hline
$c_5$ & [-0.101, -0.100] & [-0.8, -0.6] \\
\hline
\end{tabular}
\end{center}
\caption{Parameter ranges of existence. Left: creeping solitons. Right: stationary solitons.}
\label{TabCreep}
\end{table}
By changing the initial condition into a ring vortex of vorticity $m=1$, a novel class, spreading vortex soliton has been found. This soliton has exactly the same behavior as the creeping, but it is also spinning and that creates interesting ``propeller'' shape before spreading and filling the domain, see Fig. \ref{prop}. The energy of such structure starts oscillating (during this phase the beam alternatively grows up and comes back to its initial shape) and finally converges to a fixed value that takes the shape of a ring vortex. Fig. \ref{prop2} shows the evolution of the energy of a propeller soliton emphasizing this bifurcation: the energy oscillates and seems to start converging but then a perturbation makes the energy diverge and increase because of the spreading behavior of the solution. Table \ref{TabCreep2} shows a comparison between the propeller and the ring spinning type. If one compares the energy between the two, it seems like a competition between the increasing behavior of the creeping after 50 s. together with the spinning behavior of the ring for the first 50 s., when the energy wants to remain bounded see Fig. \ref{prop2}.

\begin{table}
\begin{center}
\begin{tabular}{|c|cc|}
\hline Parameters & Propeller  & Ring spinning \\
\hline
$b_1$ & [0.08, 0.14] & [0.185, 0.235] \\
\hline
$\epsilon$ & [-0.19, -0.09] & [-0.3, 0.2]  \\
\hline
$b_3$ & [-1.3, -1.23] & [-1.22, -1.11]  \\
\hline
$b_5$ & [0.3, 0.36] &  not found \\
\hline
$c_5$ & [-0.1, -0.06] & not found \\
\hline
\end{tabular}
\end{center}
\caption{Parameter ranges. Left: spreading spinning solitons. Right: ring spinning solitons.}
\label{TabCreep2}
\end{table}

%%%%%%%%%%%%%%%%%%%%%%%%%%%%%%%%%%%%%%%%%%%%%%%%
\section{Conclusion}
In conclusion, we have found new parameter ranges for the fascinating (2+1)D structures in dissipative media of the complex cubic-quintic Ginzburg-Landau equation. All classes  have been categorized by focusing on their energy, shape and phase evolution. Regions of existence of these types of soliton have been calculated in the five dimensional parameter space given by the physical parameters of the equation, plus two degrees of freedom based on the two initial conditions used, and two degrees of freedom based on vorticities $m=1, \, m=2$.  Whereas non spinning structures have mainly lead to stable stationary solitons, due to bifurcations spinning structures, symmetric or asymmetric revealed novel  solitons. By changing the initial conditions, and increasing the vorticity, allowed us to observe for the first time new types of solutions such as spinning ``bean-shaped'' solitons and creeping-vortex propeller solitons. Since the parameter space is so large this research can inspire new work to find other parameter regimes, and maybe other structures. 

The investigation involved extremely intensive computations using the existing 256-node ZEUS Linux cluster at Embry-Riddle, and it addressed the specific topics: 1) finding novel two
dimensional solitons 2) exploring the parameter
space for which they exist as stable structures, and explaining the stability loss numerically via bifurcations,
and 3) finding new parameter ranges that would yield as yet unknown types of solitons.

Also, given the generality of the
theoretical framework developed in this project, it provides a platform for the detailed modeling of
3D solitons, in the form of optical bullets. They depict a confined spatiotemporal soliton in which
the balance between the focusing nonlinearity and the beam-spreading while propagates through
media provides the shape of a bullet with potential use in optical processing of information. For the future we plan to expand our findings to the the 3+1 dimensional space, where we hope to find other types as well.
%%%%%%%%%%%%%%%%%%%%%%%%%%%%%%%%%%%%%%%%%%%%%%%%%%%%%%%%
\newpage
\bibliographystyle{plain}
\bibliography{bibliography}
\addcontentsline{toc}{section}{References}
%%%%%%%%%%%%%%%%%%%%%%%%%%%%%%%%%%%%%%%%%%%%%%%%%%%%%%%%%
\end{document}